\documentclass[]{aastex631}

\usepackage{mathtools}
\usepackage{rotating}
\usepackage{booktabs}

\def\OIIg{\ifmmode {{\rm [OII]\lambda\lambda 3726,3729}}\else
                ${\rm [OII]\lambda\lambda3726,3729}$\fi}

\newcommand{\NGDEEPA}{{NGDEEP-NISS1}}
\newcommand{\NGDEEP}{{NGDEEP-NISS}}

\def\SIIg{\ifmmode {{\rm [SII]\ \lambda\lambda6718,6733}}\else
                ${\rm [SII]\ \lambda\lambda6718,6733}$\fi}
                
\def\SII{\ifmmode {{\rm [SII]\ \lambda}}\else
                ${\rm [SII]\ \lambda}$\fi}

\def\Hag{\ifmmode {{\rm H\alpha\ \lambda6565+[NII]\lambda\lambda6550,6585}}\else
                ${\rm H\alpha\ \lambda6565+[NII]\lambda\lambda6550,6585}$\fi}

\def\NII{\ifmmode {{\rm [NII]\ \lambda}}\else
                ${\rm [NII]\ \lambda}$\fi}      

\def\NIIg{\ifmmode {{\rm [NII]\ \lambda\lambda6550,6585}}\else
                ${\rm [NII]\ \lambda\lambda6550,6585}$\fi}
         
\def\Ha{\ifmmode {{\rm H\alpha\ \lambda6565}}\else
                 ${\rm H\alpha\ \lambda6565}$\fi}
                                
\def\Hg{\ifmmode {{\rm H\gamma\ \lambda4342}}\else
                 ${\rm H\gamma\ \lambda4342}$\fi}
                                                     
\def\OIIIg{\ifmmode {{\rm [OIII]\ \lambda\lambda4960\AA,5008\AA}}\else
                ${\rm [OIII]\ \lambda\lambda4960,5008}$\fi}

\def\Hb{\ifmmode {{\rm H\beta\ \lambda4863}}\else
                ${\rm H\beta\ \lambda4863}$\fi}

\def\OIII{\ifmmode {{\rm [OIII]\ \lambda}}\else
                ${\rm [OIII]\ \lambda}$\fi}        
\def\AV{\ifmmode {{\rm A_V}}\else
                ${\rm A_V}$\fi}

\begin{document}

\title{The Next Generation Deep Extragalactic Exploratory Public Near-Infrared Slitless Survey Epoch 1 (NGDEEP-NISS1): Extra-Galactic Star-formation and Active Galactic Nuclei at 0.5 $<$ z $<$ 3.6}

\author[0000-0003-3382-5941]{Nor Pirzkal}
\affiliation{ESA/AURA Space Telescope Science Institute\\
3700 San Martin Dr. \\
Baltimore, MD, 212129}

\author[0000-0003-2283-2185]{Barry Rothberg}
\affiliation{U.S. Naval Observatory, 3450 Massachusetts Avenue NW, Washington, DC 20392, USA}
\affiliation{Department of Physics and Astronomy, George Mason University, 4400 University Drive, MSN 3F3, Fairfax, VA 22030, USA}
\email{npirzkal@stsci.edu}

\author[0000-0001-7503-8482]{Casey Papovich}
\affiliation{Department of Physics and Astronomy, Texas A\&M University, College Station, TX, 77843-4242 USA}
\affiliation{George P.\ and Cynthia Woods Mitchell Institute for Fundamental Physics and Astronomy, Texas A\&M University, College Station, TX, 77843-4242 USA}

\author[0000-0001-9495-7759]{Lu Shen}
\affiliation{Department of Physics and Astronomy, Texas A\&M University, College Station, TX, 77843-4242 USA}
\affiliation{George P.\ and Cynthia Woods Mitchell Institute for Fundamental Physics and Astronomy, Texas A\&M University, College Station, TX, 77843-4242 USA}

\author[0000-0002-9393-6507]{Gene C. K. Leung}
\affiliation{Department of Astronomy, The University of Texas at Austin, Austin, TX, USA}

\author[0000-0002-9921-9218]{Micaela B. Bagley}
\affiliation{Department of Astronomy, The University of Texas at
  Austin, Austin, TX, USA}

\author[0000-0001-8519-1130]{Steven L. Finkelstein}
\affiliation{Department of Astronomy, The University of Texas at Austin, Austin, TX, USA}

\author[0000-0002-8163-0172]{Brittany N. Vanderhoof}
\affiliation{Space Telescope Science Institute, 3700 San Martin Drive, Baltimore, MD 21218, USA}

\author[0000-0003-3130-5643]{Jennifer M. Lotz}
\affiliation{Gemini Observatory/NSF's National Optical-Infrared Astronomy Research Laboratory, 950 N. Cherry Ave., Tucson, AZ 85719, USA}

\author[0000-0002-6610-2048]{Anton M. Koekemoer}
\affiliation{Space Telescope Science Institute, 3700 San Martin Drive, Baltimore, MD 21218, USA}

\author[0000-0001-6145-5090]{Nimish P. Hathi}
\affiliation{Space Telescope Science Institute, 3700 San Martin Drive, Baltimore, MD 21218, USA}

\author[0000-0001-8551-071X]{Yingjie Cheng}
\affiliation{University of Massachusetts Amherst, 710 North Pleasant Street, Amherst, MA 01003-9305, USA}

\author[0000-0001-7151-009X]{Nikko J. Cleri}
\affiliation{Department of Physics and Astronomy, Texas A\&M University, College Station, TX, 77843-4242 USA}

\author[0000-0003-3466-035X]{{L. Y. Aaron} {Yung}}
\affiliation{Space Telescope Science Institute, 3700 San Martin Drive, Baltimore, MD 21218, USA}
\affiliation{George P.\ and Cynthia Woods Mitchell Institute for Fundamental Physics and Astronomy, Texas A\&M University, College Station, TX, 77843-4242 USA}

\author[0000-0001-8534-7502]{Bren E. Backhaus}
\affil{Department of Physics, 196A Auditorium Road, Unit 3046, University of Connecticut, Storrs, CT 06269, USA}

\author[0000-0003-2098-9568]{Jonathan P. Gardner}
\affiliation{Astrophysics Science Division, NASA Goddard Space Flight Center, 8800 Greenbelt Rd, Greenbelt, MD 20771, USA}

\author[0000-0003-4528-5639]{Pablo G. P\'erez-Gonz\'alez}
\affiliation{Centro de Astrobiolog\'{\i}a (CAB), CSIC-INTA, Ctra. de Ajalvir km 4, Torrej\'on de Ardoz, E-28850, Madrid, Spain}

\author[0000-0001-7113-2738]{Henry C. Ferguson}
\affiliation{Space Telescope Science Institute, 3700 San Martin Drive, Baltimore, MD 21218, USA}

\author[0000-0001-9440-8872]{Norman A. Grogin}
\affiliation{Space Telescope Science Institute, 3700 San Martin Drive, Baltimore, MD 21218, USA}

\author[0000-0002-7547-3385]{Jasleen Matharu}
\affiliation{Cosmic Dawn Center, Niels Bohr Institute, University of Copenhagen, R\aa dmandsgade 62, 2200 Copenhagen, Denmark\\}

\author[0000-0002-5269-6527]{Swara Ravindranath}
\affiliation{Space Telescope Science Institute, 3700 San Martin Drive, Baltimore, MD 21218, USA}

\author[0000-0003-0894-1588]{Russell Ryan}
\affiliation{Space Telescope Science Institute, 3700 San Martin Drive, Baltimore, MD 21218, USA}

\author[0000-0002-4153-053X]{Danielle A. Berg}
\affiliation{Department of Astronomy, The University of Texas at Austin, Austin, TX, USA}

\author[0000-0002-0930-6466]{Caitlin M. Casey}
\affiliation{Department of Astronomy, The University of Texas at Austin, Austin, TX, USA}
\affiliation{Cosmic Dawn Center (DAWN), Denmark}

\author[0000-0001-9875-8263]{Marco Castellano}
\affiliation{INAF - Osservatorio Astronomico di Roma, via di Frascati 33, 00078 Monte Porzio Catone, Italy}

\author[0000-0002-0786-7307]{\'{O}scar A. Ch\'{a}vez Ortiz}
\affiliation{Department of Astronomy, The University of Texas at Austin, Austin, TX, USA}

\author[0000-0003-4922-0613]{Katherine Chworowsky}
\affiliation{Department of Astronomy, The University of Texas at Austin, Austin, TX, USA}
\altaffiliation{NSF Graduate Fellow}

\author[0000-0001-5414-5131]{Mark Dickinson}\affiliation{NSF's National Optical-Infrared Astronomy Research Laboratory, 950 N. Cherry Ave., Tucson, AZ 85719, USA}

\author[0000-0002-6748-6821]{Rachel S. Somerville}
\affiliation{Center for Computational Astrophysics, Flatiron Institute, 162 5th Avenue, New York, NY, 10010, USA}

\author[0000-0002-1803-794X]{Isabella G. Cox}
\affiliation{Laboratory for Multiwavelength Astrophysics, School of Physics and Astronomy, Rochester Institute of Technology, 84 Lomb Memorial Drive, Rochester, NY 14623, USA}

\author[0000-0003-2842-9434]{Romeel Dav\'e}
\affiliation{Institute for Astronomy, University of Edinburgh, Royal Observatory, Edinburgh, EH9 3HJ, UK}
\affiliation{University of the Western Cape, Bellville, Cape Town 7535, South Africa}

\author[0000-0001-8047-8351]{Kelcey Davis}
\altaffiliation{NSF Graduate Research Fellow}
\affiliation{Department of Physics, 196 Auditorium Road, Unit 3046, University of Connecticut, Storrs, CT 06269, USA}

\author[0000-0001-8489-2349]{Vicente Estrada-Carpenter}\affiliation{Department of Astronomy \& Physics, Saint Mary's University, 923 Robie Street, Halifax, NS, B3H 3C3, Canada}

\author[0000-0003-3820-2823]{Adriano Fontana}
\affiliation{INAF Osservatorio Astronomico di Roma, Via Frascati 33, 00078 Monteporzio Catone, Rome, Italy}

\author[0000-0001-7201-5066]{Seiji Fujimoto}
\altaffiliation{Hubble Fellow}
\affiliation{Department of Astronomy, The University of Texas at Austin, Austin, TX, USA}

\author[0000-0002-7831-8751]{Mauro Giavalisco}
\affiliation{Department of Astronomy, University of Massachusetts, Amherst, MA 01003, USA}
\
\author[0000-0002-5688-0663]{Andrea Grazian}
\affil{INAF--Osservatorio Astronomico di Padova,
Vicolo dell'Osservatorio 5, I-35122, Padova, Italy\\}

\author[0000-0001-6251-4988]{Taylor A. Hutchison}
\altaffiliation{NASA Postdoctoral Fellow}
\affiliation{Astrophysics Science Division, NASA Goddard Space Flight Center, 8800 Greenbelt Rd, Greenbelt, MD 20771, USA}

\author[0000-0002-6790-5125]{Anne E. Jaskot}
\affiliation{Department of Astronomy, Williams College, Williamstown, MA, USA}

\author[0000-0003-1187-4240]{Intae Jung}
\affiliation{Space Telescope Science Institute, 3700 San Martin Drive, Baltimore, MD 21218, USA}

\author[0000-0001-9187-3605]{Jeyhan S. Kartaltepe}
\affiliation{Laboratory for Multiwavelength Astrophysics, School of Physics and Astronomy, Rochester Institute of Technology, 84 Lomb Memorial Drive, Rochester, NY 14623, USA}\

\author[0000-0001-8152-3943]{Lisa J. Kewley}
\affiliation{Center for Astrophysics | Harvard \& Smithsonian, 60 Garden Street, Cambridge, MA 02138, USA}

\author[0000-0002-5537-8110]{Allison Kirkpatrick}
\affiliation{Department of Physics and Astronomy, University of Kansas, Lawrence, KS 66045, USA}

\author[0000-0002-8360-3880]{Dale D. Kocevski}
\affiliation{Department of Physics and Astronomy, Colby College, Waterville, ME 04901, USA}

\author[0000-0003-2366-8858]{Rebecca L. Larson}
\altaffiliation{NSF Graduate Fellow}
\affiliation{Department of Astronomy, The University of Texas at Austin, Austin, TX, USA}

\author[0000-0002-5554-8896]{Priyamvada Natarajan}
\affiliation{Department of Astronomy, Yale University, 52 Hillhouse Avenue, New Haven, CT 06511, USA}
\affiliation{Department of Physics, Yale University, P.O. Box 208121, New Haven, CT 06520, USA}
\affiliation{Black Hole Initiative at Harvard University, 20 Garden Street, Cambridge, MA 02138, USA}

\author[0000-0001-8940-6768]{Laura Pentericci}
\affiliation{INAF - Osservatorio Astronomico di Roma, via di Frascati 33, 00078 Monte Porzio Catone, Italy}

\author[0000-0002-6386-7299]{Raymond C. Simons}
\affiliation{Department of Physics, 196 Auditorium Road, Unit 3046, University of Connecticut, Storrs, CT 06269, USA}

\author[0000-0002-4226-304X]{Gregory F. Snyder}
\affiliation{Space Telescope Science Institute, 3700 San Martin Drive, Baltimore, MD 21218, USA}

\author[0000-0002-1410-0470]{Jonathan R. Trump}
\affiliation{Department of Physics, 196 Auditorium Road, Unit 3046, University of Connecticut, Storrs, CT 06269, USA}

\and

\author[0000-0003-3903-6935]{Stephen M.~Wilkins} %
\affiliation{Astronomy Centre, University of Sussex, Falmer, Brighton BN1 9QH, UK}
\affiliation{Institute of Space Sciences and Astronomy, University of Malta, Msida MSD 2080, Malta}



\begin{abstract}
The Next Generation Deep Extragalactic Exploratory Public (NGDEEP) survey program was designed specifically to include Near Infrared Slitless Spectroscopic observations (\NGDEEP) to detect multiple emission lines in as many galaxies as possible and across a wide redshift range using the Near Infrared Imager and Slitless Spectrograph (NIRISS). We present early results obtained from the the first set of observations (Epoch 1,  50$\%$ of the allocated orbits) of this program (\NGDEEPA).  Using a set of independently developed calibration files designed to deal with a complex combination of overlapping spectra, multiple position angles, and multiple cross filters and grisms, in conjunction with a robust and proven algorithm for quantifying contamination from overlapping dispersed spectra, \NGDEEPA\ has achieved a 3$\sigma$ sensitivity limit of 2 $\times$ 10$^{-18}$ erg/s/cm$^2$. We demonstrate the power of deep wide field slitless spectroscopy (WFSS) to characterize the star-formation rates, and metallicity ([OIII]/H$\beta$), and dust content, of galaxies at $1<z<3.5$.  The latter showing intriguing initial results on the applicability and assumptions made regarding the use of Case B recombination. 
 Further, we identify the presence of active galactic nuclei (AGN) and infer the mass of their supermassive black holes (SMBHs) using broadened restframe MgII and H$\beta$ emission lines. The spectroscopic results are then compared with the physical properties of galaxies extrapolated from fitting spectral energy distribution (SED) models to photometry alone.  The results clearly demonstrate the unique power and efficiency of WFSS at near-infrared wavelengths over other methods to determine the properties of galaxies across a broad range of redshifts.
\end{abstract}


\section{Introduction} \label{sec:intro}
A critical component to maximizing surveys of emission line of galaxies is the balance among  efficiency, depth, and resolving targets sufficiently to fully characterize their physical properties.  Purely photometric surveys may be the most efficient method in terms of depth and sensitivity versus time required, but their accuracy in determining redshifts, dust, metallicity, stellar ages, etc. are highly dependent upon the input models used to match and fit the observed data, the methods and assumptions which go into such modeling, and a sufficient number of filters to properly sample the spectral energy distributions (SEDs) of galaxies.  As more filters are used, the efficiency of such surveys lessens.  Spectroscopy is the gold standard and should always be used to verify redshifts, and other galaxy properties.  However, spectroscopic observations are not without complications.  Extracting parameters from objects beyond the local Universe can be extremely time expensive, even with the largest apertures.  At wavelengths beyond 0.8 $\mu$m, such observations are further complicated by telluric sky emission and atmospheric absorption, as well as thermal contributions from the sky and instruments at longer near-infrared wavelengths.  Further, the nuts and bolts of standard spectroscopic observations, such as slit alignment, constructing multi-object-slit masks to efficiently maximize a survey field, maximizing multi-fiber placement, or the various complexities of integral field unit observations (field of view, wavelength coverage, spectral resolution, etc) can hinder the depth and efficiencies of such observations.  

Wide field slitless spectroscopy (WFSS) provides a balance between the limitations of purely photometric surveys and the complexities of standard types of spectroscopic observations.  Such techniques are neither new, nor novel \citep[See][for a brief review]{Pirzkal17a}.  WFSS disperses the light of every object observed in the field of view (FOV), thus eliminating the problems of slit loss, misplaced fibers, etc..  The spectral resolution (R) of WFSS is significantly less than standard spectroscopy (R$\simeq$ 10s-100 versus $>>$ few hundred) but it is sufficient to measure important characteristics beyond just redshift, such as metallicity, dust, and even detect active galactic nuclei (AGN).  One downside to WFSS is that spectral contamination can occur from overlapping spectra produced by sources in close proximity to each other, and/or from multiple orders dispersed across the detector.  However, with careful planning (e.g. observations taken with different orientations on-sky and/or using orthogonally crossed dispersers) and careful forward modeling of contamination \citep[e.g. EM2D, MAP2D, see][for details]{Pirzkal17b,Pirzkal18}, one can reduce the impacts of spectral contamination.  Moving beyond Earth's atmosphere alleviates the impact of telluric absorption and emission lines and the limitations of limited atmospheric transparency windows in the near-infrared.  Observations from the Hubble Space Telescope (HST) are still impacted from Earth glow in certain orientations \citep{Brammer14,Pirzkal20,Pirzkal17a,Simons23} and along with the James Webb Telescope \citep[JWST][]{Gardner23} may still be affected by thermal contributions from the instruments and telescope itself, but at levels significantly less than ground-based observations. 

\NGDEEPA\ (The Next Generation Deep Extragalactic Exploratory Public Near-Infrared Slitless Spectroscopic survey 1) follows a legacy of programs using space-based WFSS deep surveys of the $\approx 3' \times 3'$\ Hubble Ultra Deep Field \citep[HUDF,][]{Beckwith06}, specifically the Grism ACS Program for Extragalactic Science \citep[GRAPES,][]{Pirzkal04} the Probing Evolution And Reionization Spectroscopically \citep[PEARS,][]{Pirzkal09}, which both used the the optical (0.4$\mu$m $<$$\lambda$$<$ 1$\mu$m) Advanced Camera for Surveys (ACS),  and later the Faint Infrared Grism Survey \citep[FIGS,][]{Pirzkal17a} using the Wide Field Camera 3 (WFC3), with an emphasis on the near-infrared camera, which extended coverage to 1.6$\mu$m.  The programs cited above come in addition to WFSS programs that, while shallower, covered larger areas such as WISPS \citep[][]{Atek10}, 3DHST \citep{brammer12, Momcheva16}, and CLEAR \citep[][]{Simons23}.
Until the launch of JWST, space-based WFSS surveys at longer wavelengths were not possible.  \NGDEEP\ was designed specifically to take advantage of the capabilities of JWST's Near Infrared Imager and Slitless Spectrograph \citep[NIRISS][]{Doyon23}, which covers a wavelength range 1$\mu$m $<$ $\lambda$ $<$ 2.3$\mu$m (using the three cross filters F115W, F150W, and F200W) over a 2\arcmin.2 $\times$ 2\arcmin.2 FOV, with a spectral resolution of R$\simeq$ 150.  The goals of the program include the study of emission line galaxies at higher redshifts, detecting faint emission lines, and leveraging the wide wavelength range to detect multiple emission lines which allows for more robust scientific analysis of galaxy properties such as gas metallicity, the Balmer decrement (internal dust extinction), BPT diagrams \citep[the Baldwin, Phillips \& Terlevich star-formation versus AGN diagnostic,][]{Baldwin81}, among other parameters, for galaxies over the redshift range of 1 $<$ z $<$ 3.5. For a more in-depth overview of NGDEEP, the reader is referred to \cite{Bagley23}.  Further, the angular spatial resolution of NIRISS (0.09-0.1\arcsec), in combination with four distinct orientations (two orthogonal grisms used at two position angles on the sky) allows for \NGDEEP\ observations to be used to construct two-dimensional spatial maps of galaxies as a function of wavelength.  That is, for each resolved source in the field, one can resolve individual areas of emission beyond the nucleus and map the presence and intensity of star-formation, which helps to further constrain the evolutionary history of these systems.

Exactly half of the NGDEEP-NIRISS observations were obtained in February 2023, using two grisms but at a single orientation on the sky,  and has achieved an average sensitivity of 1.88 $\times$ 10$^{-18}$ erg/s/cm$^2$ (for a 3$\sigma$ detection, see Table \ref{tab:limits}  for more details).  This paper presents the first look at these data which we refer to as \NGDEEPA\ in this paper.  Observations targeted the HUDF at the coordinates of 03:32:38.6007, -27:46:59.83 and the first Epoch of the data was taken at a position angle (PA) of 70 degrees. All NIRISS observations were taken using the NIS readout mode. Direct imaging for NIRISS used the F115W, F150W, and F200W filters employing one single integration of 5 groups each. The NIRISS grism observations were made with the following configurations:  All observations were made with the G150R and G150C grisms;  The F115W cross filter used 3 exposures with 6 integrations per exposure and 20 groups per integration; The F150W and F200W cross filters were obtained with 1 exposure each and 8 integrations per exposure and 20 groups per integration. The total exposure time for each of the two grisms were  47Ks, 21Ks, and 16Ks, for the F115W, F150W, and F200W cross filters, respectively. The dithering pattern was dictated by the fact that NIRcam imaging was obtained in parallel and was the 3-POINT-LARGE-NIRCam pattern.
 The goals of this paper are:  1) showcase the quality of the \NGDEEPA\ data and how it compares to how pre-launch expectations; 2) show how \NGDEEPA\ emission line measurements of extinction and star-formation rates differ from the same properties obtained from fitting spectral energy distributions (SEDs) to photometry alone; and 3) explore the redshift evolution of the population of star forming galaxies at $1<z<3.5$. 

All calculations in this paper assume $H_0 = 67.3\ km\ s^{-1} Mpc^{-1}$\ and $\Omega_M = 0.315$, $\Omega_\Lambda = 0.685$ \citep{Planck15}. All magnitudes are given in the $AB$\ system \citep{Oke83}.
\\

\section{Data Reduction and Analysis} \label{sec:style}
\subsection{Pipeline and Catalogs}
We started by using the rate files STScI (Space Telescope Science Institute) Pipeline\footnote{see \url{https://jwst-docs.stsci.edu/jwst-science-calibration-pipeline-overview/stages-of-jwst-data-processing} for details} Stage 1 products to perform the bias subtraction, dark current correction, and on-the-ramp fitting to produce partially calibrated slope images. The pipeline Stage 2 {\em calwebb\_image2} was used for all data, including WFSS observations in order to populate the world coordinate system (WCS) of each image using {\em assign\_wcs}, and apply the appropriate imaging filter flat-field using {\em flat\_field}. The final pipeline processing of the observations was completed in August 2023, using the STScI pipeline version 1.10.2, CRDS version 11.17.0 and the CRDS context file {\em jwst\_1110.pmap} (these were the most up to date at the time.)
These steps were performed in order to be able to produce 1D fully calibrated spectra using an implementation of the SBE (Simulation Based Extraction) technique detailed in \citet{Pirzkal17a} from partially calibrated WFSS observations in units of DN/s.

Part of the process of selecting and extracting WFSS data requires the presence of a field image.  This image (or images) is (are) used to match the dispersed light to its source. For the \NGDEEPA\ data, four sets of images were used.  First, a catalog of source candidates to extract was generated from SExtractor \citep{Bertin96} using a HST F160W master mosaic. This master mosaic and its segmentation maps, also used by \citet{Lu23}, incorporated the latest reduction of the HST observations from the Cosmic Assembly Near-IR Deep Extragalactic Legacy Survey \citep[CANDELS,][]{Koekemoer11,Grogin11,Finkelstein22}. It was used to provide a segmentation map that was astrometrically corrected to match the GAIA Data Release 3 (DR3) catalog.  The choice of the F160W mosaic, with a $5\sigma$\ sensitivity limit of $mag_{AB}=27.5$, to generate this early segmentation map could, in principle, miss some of the faintest red objects.  Once the completed NGDEEP observations have been reduced, future papers will examine the combined direct imaging from both epochs and compare with the F160W HST mosaic in order to fully characterize potential missed objects. See \cite{Koekemoer11} and Section 2.2 of \cite{Leung23} for more details regarding the creation of this F160W mosaic creation, as well as mosaics in additional bands using other existing HST/ACS and HST/WFC3 data. The depth of this mosaic is otherwise several magnitudes deeper than the sources discussed in this paper and allows for the estimation of the spectral contamination from other sources that are more than 4 magnitudes fainter. Next, a second set of mosaiced fields were created from the \NGDEEPA\ images in each of the F115W, F150W, and F200W filters. These imaging observations were taken during the same pointing of the JWST WFSS observations and represent a 1:1 match between objects and their dispersed light.  These NIRISS F115W, F150W, and F200W mosaics were created using the JWST Level 3 pipeline.  However, these were {\it not} corrected to match the astrometry from the GAIA DR3 catalog, as this would break the relative alignment between the \NGDEEP\ imaging and \NGDEEP\ WFSS data. Instead, sources were detected in each of the individual (F115W, F150W, and F200W) \NGDEEP\ mosaics.  An affine transformation (shift and rotation) was then derived between each mosaic and the HST F160W  master mosaic. Thus, the coordinates of any of the pixels in the master F160W mosaic could be accurately (to within $<0.1$\ pixel) mapped onto the individual \NGDEEP\ mosaics, and thus by extension, onto the individual WFSS observations.

\subsection{Photometry}
NIRISS WFSS were used in conjunction with photometry of the HUDF from HST/ACS (F435W, F606W, F775W, F814W, and F850LP filters), HST/WFC3-IR (F105W, F125W, F140W, and F160W filters), and JWST/NIRISS imaging (F115W, F150W, and F200W filters), for a total of 12 bands which span a range of observed wavelengths from 0.4 to 2.5$\mu$m.  The analysis and methods for constructing the mosaic images and measuring the photometric data can be found in \cite{Leung23}.  The photometric data were used exclusively to calculate stellar masses, star-formation histories, and SEDs.  They were also used to determine metallicities, extinction values and dust corrections, and redshifts.  These values were then compared to those determined directly from WFSS data.  The masses derived from photometry were also used to correct [NII] contamination from \Ha\ for generating BPT diagrams.  

\subsection{WFSS Extraction and Calibration} \label{WFSS Extraction and Calibration}
Following the method described in \citep{Pirzkal17b}, the dispersed WFSS observations were fully simulated using the object pixel coordinates from the master mosaic and the pixel level photometry of the individual F115W, F150W, and F200W mosaics: For every galaxy, a defined smooth spectral energy distribution (SED) was generated for each imaging pixel of the galaxy, which could then be dispersed numerically to produce an accurate and smooth simulation of the source in each individual \NGDEEP\ WFSS dataset.
The initial simulations using the official STScI NIRISS WFSS calibration files (based on commissioning data and some Cycle 1 observations) showed large offsets (as much as 2 pixels, or $\approx 90\AA$\ in the dispersion direction) between the expected and measured positions of dispersed spectral traces in the observations. These errors were too large to quantitatively and rigorously correct the contamination from overlapping spectra or to produce accurate observed wavelengths of emission lines when using different grisms and cross filters. Further, these errors create spurious features which could be misidentified  or misinterpreted as real physical results.

Since \NGDEEP\ combines multiple spectra, obtained using two cross filters, different grisms, and two orientations on the sky, the official STScI calibration products, with a precision on the order of 1 to 2 pixels, proved inadequate for our needs. Therefore, a new set of calibration files for the NIRISS WFSS mode were derived for the F115W, F150W, and F200W imaging cross filters with the GR150R and G150C grisms to bring the calibration uncertainties down to about 0.2 pixels, both in terms of trace location and shape as well as wavelength calibration. This work followed the methodology used to calibrate the HST/WFC3 WFSS mode \citep[See][for details]{Pirzkal16,Pirzkal17c}. Appendix \ref{calib} provides quantitative comparisons regarding the accuracy achieved from these calibrations, and the improvement in accuracy reached by the \NGDEEP\ calibration when compared to the STScI NIRISS WFSS calibration. {\it All} of the work described in this paper, including simulations and extractions, used the improved \NGDEEP\ calibrations.  These calibrations will be made publicly available as a service to the community.

\begin{figure*}
\center
\includegraphics[width=6.5in]{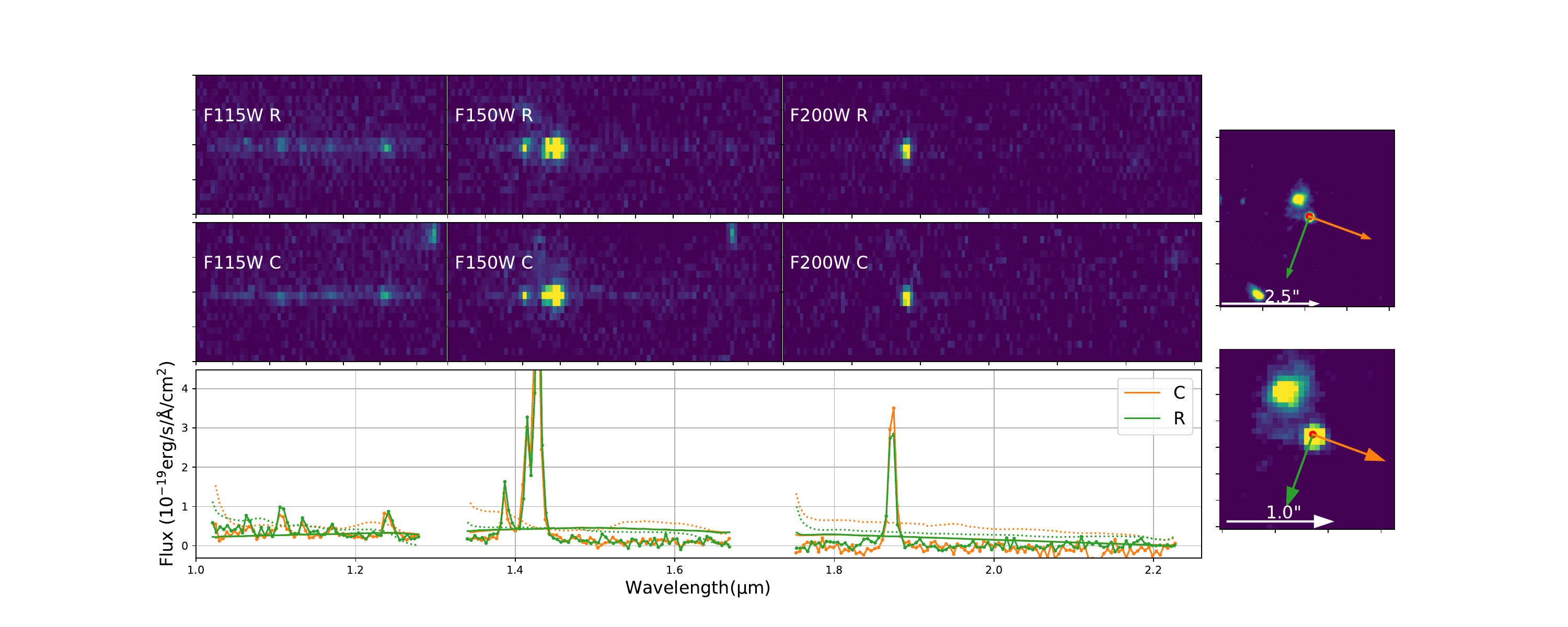}
\caption{An example spectrum of one of the \NGDEEPA\ sources (ID $\#$33273). The rightmost panels show two views of the field image, the top is a larger field of view 5$\arcsec$ $\times$ 5$\arcsec$ arcseconds in size, the bottom is a zoom-in 2$\arcsec$\ $\times$ 2$\arcsec$\ in size.
  The color arrows show the directions of the light dispersed by the G150C (orange) and G150R (green) grisms. Prominent emission lines are seen with the F115W, F150W, and F200W cross filters (Rows 1 and 2: left, center, and right, respectively) used in conjunction with the GR150C (Row 1 and in orange in the bottom plot) and GR150R (Row 2 and in green in the bottom plot) NIRISS grisms. This source has a magnitude of $m_{AB, F160W}=24.8$\ and is at z=1.85. The dotted lines show the contamination estimates which were subtracted from the observed spectra. The initial models, estimated using photometry, are also plotted (solid smooth lines). No continuum was subtracted from these spectra.\label{2Dexample}}
\end{figure*}

Similarly to what was done in \citet{Pirzkal17a}, full 2D simulations were used to estimate and subtract the background from individual WFSS observations, and to provide quantitative estimates of the WFSS contamination resulting from overlapping dispersed spectra. First, individual rectified 2D spectra were generated from each observation.  Next, these were subsequently combined to produce deep, rectified 2D spectra, from which fully calibrated 1D spectra  are generated using an optimal extraction method closely based on \citep{horne86}. No galactic correction was applied to the extracted spectra to account for the foreground dust of the Milky Way Galaxy in the direction of the HUDF because the reddening is $ E(B-V) \approx 0.007 \pm 0.0004$ \citep{Schlafly11}. Then, for each source, 6 individual spectra were generated using the GR150R and GR150C grisms, combined with the F115W, F150W, and F200W cross filters. GR150R and GR150C are identical grisms but oriented perpendicularly. Figures \ref{2Dexample} and \ref{2Dshift} show examples of 2D rectified spectra, as well as the final, calibrated 1D spectra.  Multiple emission lines such as \OIIg, \Hb, \Hag, \OIIIg, are clearly detected in both figures.  Spectra of similar quality have been extracted out to z $\sim$ 3.

\subsection{Emission lines detection and fitting} \label{sec:floats}
 We extracted the spectra of sources brighter than $m_{AB,F160W}=26$, for a total of 825 sources. The final 1D extracted spectra were visually inspected and a set of 144 sources with easily identifiable emission lines (\Hag + \SIIg, and/or \OIIIg+\Hb ) were manually selected for analysis in this paper. These sources have magnitudes of $21.24 < m_{AB,F160W} < 25.96$.    The identification and initial fitting of these emission line in the 1D spectra allowed for the determination of accurate spectroscopic redshifts for each of the sources.  The line selection was limited to manual identification of  \Hag+\SIIg, and/or \OIIIg+\Hb\ lines and their observed wavelengths in the extracted spectra. The requirement for this was that the emission lines should be identifiable in {\it both} the GR150R and GR150C grisms.  Once a redshift for each source was set, the lines were subsequently fit, and the fluxes of emission lines estimated. 

\begin{figure*}
\center
\includegraphics[width=6.5in]{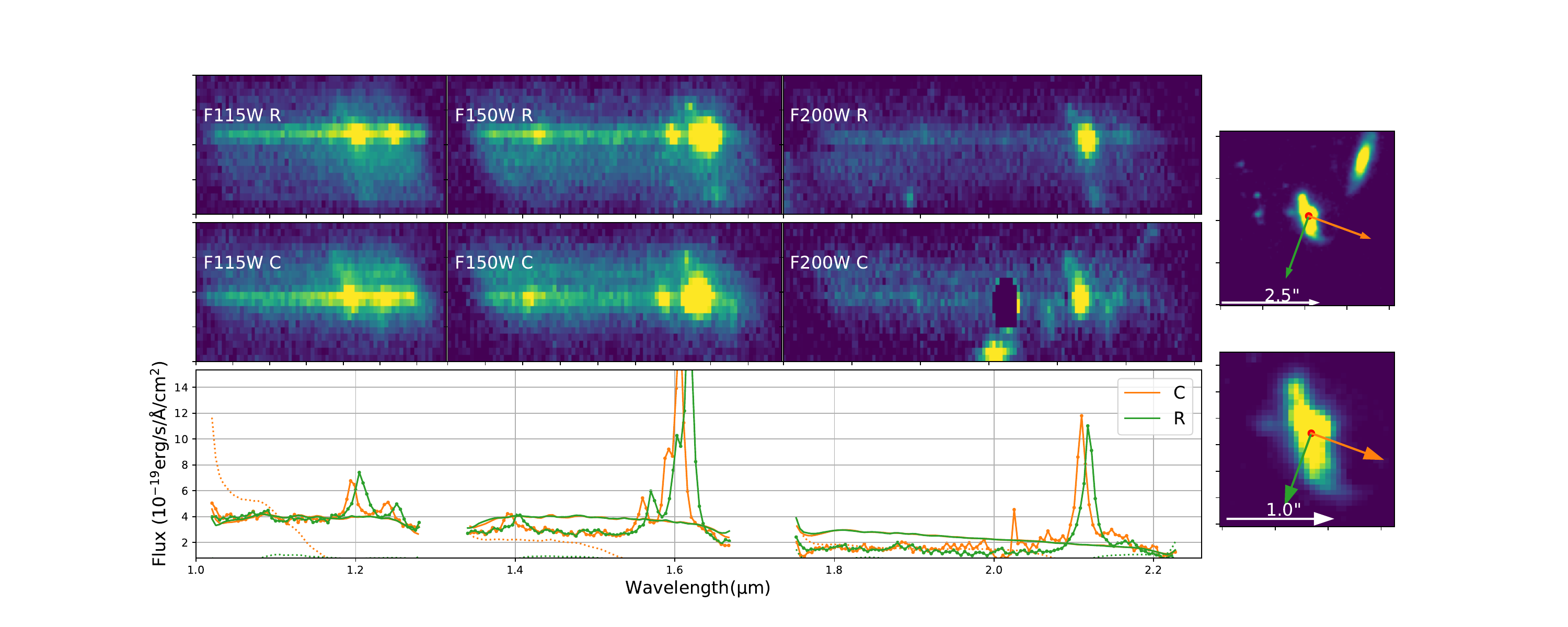}

\caption{Similar to Figure \ref{2Dexample} (colors and lines have the same meaning), however, this spectra of an \NGDEEPA\ target (ID$\#$35875) demonstrates the offset between the orthogonal grisms which is due the presence of off-nuclear, spatially resolved star-formation.  This offset produces different {\it observed} wavelengths being assigned to emission lines in the R and C grisms. Simply combining or averaging the two orthogonal grisms without taking this into account could lead to spurious scientific conclusions. The host galaxy is a  $m_{AB, F160W}=22.48$\ at z=2.23.  The blackened pixels in the bottom dispersed figure represents data masked due to contamination which cannot be corrected. \label{2Dshift}}
\end{figure*}

It is important to note that the {\it initial line identifications} used to determine redshifts {\it are only a starting point}. The excellent spatial resolution of NIRISS WFSS allows for resolved spectra to be measured spatially across resolved or partially-resolved sources.  This allows one to detect {\it extra-nuclear} emission lines resulting from star-forming regions or outflows potentially related to AGN.  As demonstrated in \citet{Pirzkal18} for HST/WFC3 WFSS, such shifts must be carefully accounted for, including the impact from self-contamination (e.g. multiple emission line sources along the dispersion axis, with each source dispersing light on top of each other).  That is, each position angle should be extracted and 1D spectra generated separately.  In the case of NIRISS, in addition to each position angle, each orthogonal grism must be treated individually. This can be used to construct spatially resolved 2D spectra of targets when obtaining spectra at multiple position angles and/or orthogonal grisms. Figure \ref{2Dshift} shows an example of a spatially resolved target and the resulting G150C and G150R 2D dispersed spectra, along with the 1D spectrum extracted from each grism.  Also shown are postage stamp images of the targets at two different fields of view.  Overplotted on the postage stamp images are lines showing the dispersion direction for each grism.  Since the position of the source of the dispersed light is different in each grism (or contains multiple sources), the resulting spectra will produce an {\it apparent} wavelength shift.  Further, if an object is resolved, or partially resolved and has any ellipticity, then any differences in the full width at half maximum (FWHM) of the source between the orthogonal grisms will produce a significant shift. 

\begin{figure*}
\center
\includegraphics[width=6.5in]{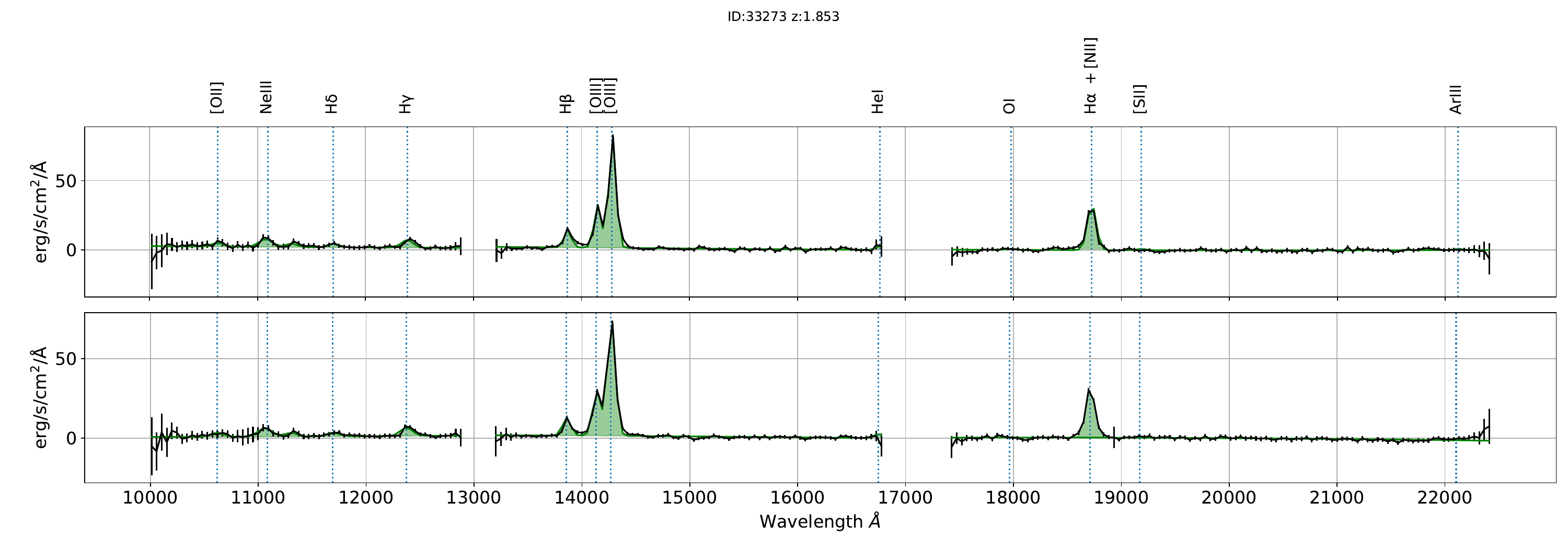}
\caption{Fits to the continuum and emission lines of the source shown in Figure \ref{2Dexample}. The measured spectra are shown in black while we show the emission line fits in green. Prominent emission lines are indicated using vertical lines.    \label{FITsample1}}
\end{figure*}

\begin{figure*}
\center
\includegraphics[width=6.5in]{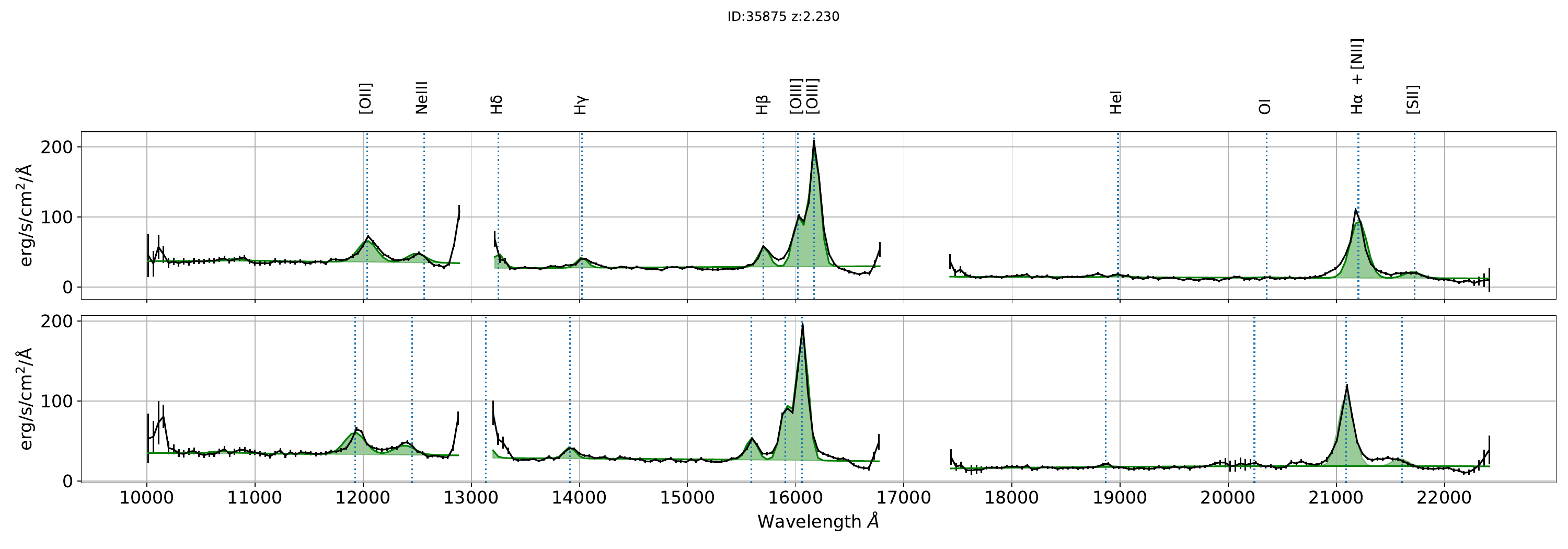}
\caption{ Similar to Figure \ref{FITsample1}, (colors and lines have the same meaning). This figure shows} the continuum and emission lines of the source shown in Figure \ref{2Dshift}.   \label{FITsample2}
\end{figure*}

\begin{figure*}
\center
\includegraphics[width=7.5in]{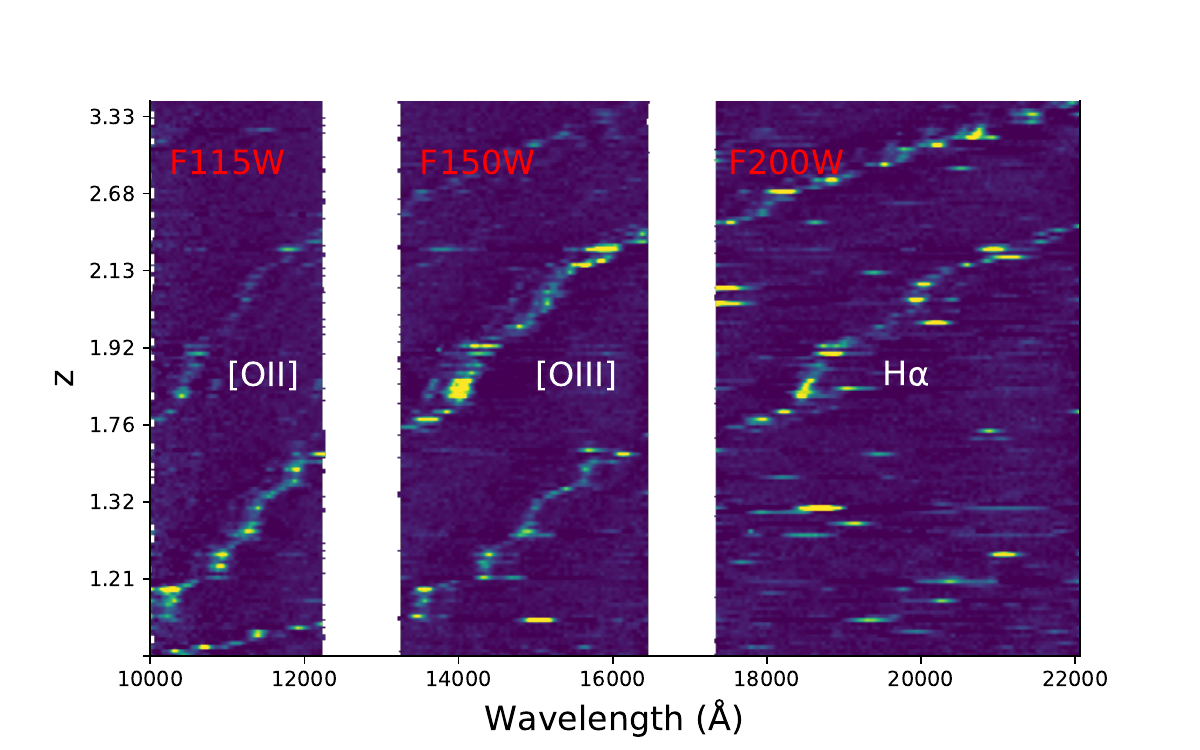}
\caption{Stacked 1D \NGDEEPA\ spectra, ordered as a function of redshift. The [OII], [OIII], and ${\rm H\alpha}$ lines are indicated.
Lines other than [OII], [OIII], and ${\rm H\alpha}$ are present in this figure.  A list of these lines are presented in Table \ref{tab:slines} and will be addressed in more detail in future papers.}
\label{AllEMM}
\end{figure*}

Therefore, all measurements of the emission line fluxes were made by fitting {\it all} emission lines simultaneously in {\it each} of the 6 available spectra (two orthogonal grisms for each of the three filters).  This was done using a single flux value for each emission line, but allowing the FWHM of the lines, and the continuum to vary in each of the 6 spectra.  The continuum emission was modelled as a second order polynomial. Finally, in order to account for shifts between the GR150R and GR150C spectra caused by extra-nuclear located emission line regions, a linear offset (in wavelength) between the GR150R and GR150C spectra were included.  The GR150R observed wavelengths were used as the fiducial  spectra. The selection of which grism to use is arbitrary, and the G150C could also be used as the fiducial spectra without impacting the final results. A custom line fitting code to perform these measurements were developed using Bilby \citep{bilby} with the Dynesty \citep{dynesty} sampler, which are an implementation of Markov-Chain Monte Carlo (MCMC) analysis. This approach leveraged all of the available information which allowed for an estimate of the emission line fluxes' posteriors, with reliable confidence regions, and realistic error estimates. For the purposes of this paper, all emission lines selected were those with  SNR$>2$.  Figures \ref{FITsample1} and  \ref{FITsample2} show the final 1D spectra generated from the two orthogonal grisms plotted in each of Figures \ref{2Dexample} and \ref{2Dshift}.  Figure \ref{AllEMM} shows a stacked image of all the emission lines spectra (SNR$>2$) discussed in this paper, ordered by redshift.

\section{Results}\label{Results}
The results presented here represent the first set of data obtained with \NGDEEPA.  The goal of this paper is to present first results, not a comprehensive or in-depth analysis from the data. But more importantly, it is to demonstrate the capabilities of carefully calibrated WFSS data from NIRISS and what physical parameters can be determined from these data. Results are also presented from combining \NGDEEPA\ data with ancillary and complimentary photometric data.  The NIRISS WFSS data are used exclusively to calculate star-formation rates (SFR), extinction values and dust corrections using the Balmer decrement \citep[e.g.][]{balmerdec31,BakerMenzel38a,millermathews72}, and of course, redshift determinations. The quality of the \NGDEEPA\ data is such, that AGN candidates have been identified via the BPT diagram and another z $\sim$ 3.18 candidate identified from broadened emission lines for which an estimate of the black hole mass has been made.  In each sub-section below we detail the various physical parameters which can be extracted from the \NGDEEPA\  data and serve as an example of the power of deep NIRISS/WFSS observations with JWST.  In all the work shown in this paper, asymmetrical error bars were estimated by using the full posterior distributions of the measured emission line fluxes and the full posteriors were furthermore propagated throughout any computation made in order to derive realistic error estimates.

\subsection{Emission Lines Fluxes}
\begin{figure*}
\center
\includegraphics[width=6.5in]{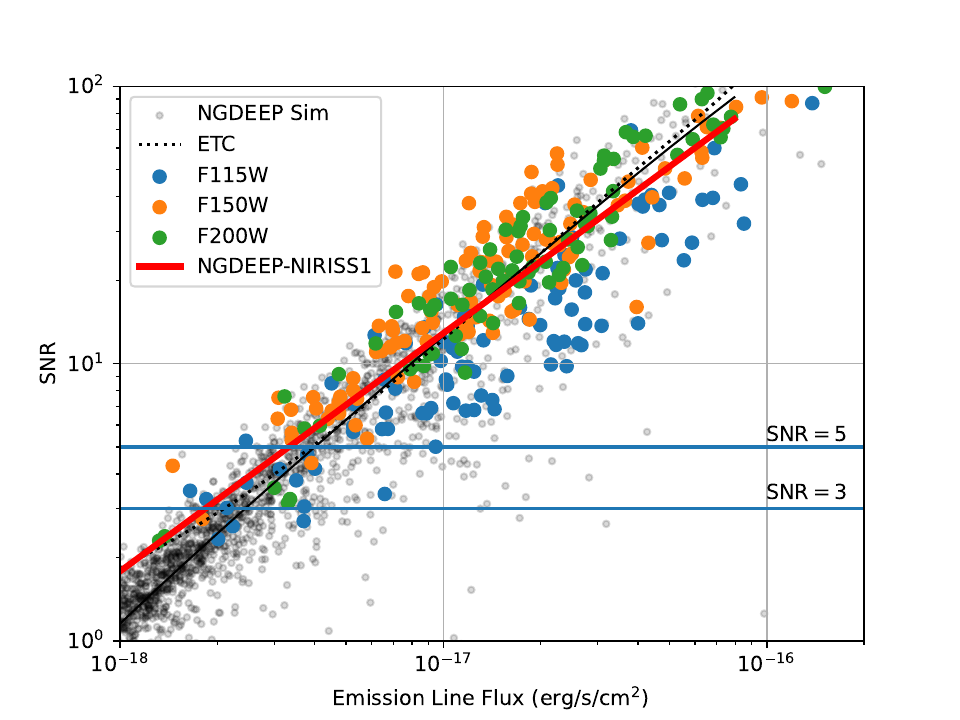}
\caption{Signal to noise (SNR) of the \NGDEEPA\ emission lines as a function of emission line flux. Shown are emission lines detected using the F115W (blue), F150W (orange), and F200W (green) cross filters, as well as a polynomial fit to the data. The F150W and F200W data achieve similar sensitivities, while the F115W data are somewhat noisier. Also plotted are the signal to noise ratios of our pre-launch simulated data (black points), the fit to this data (red line).  These are then compared with the estimates using the JWST ETC 2.0 (black dotted line).   \label{SNR}}
\end{figure*}

\begin{figure*}
\center
\includegraphics[width=6.5in]{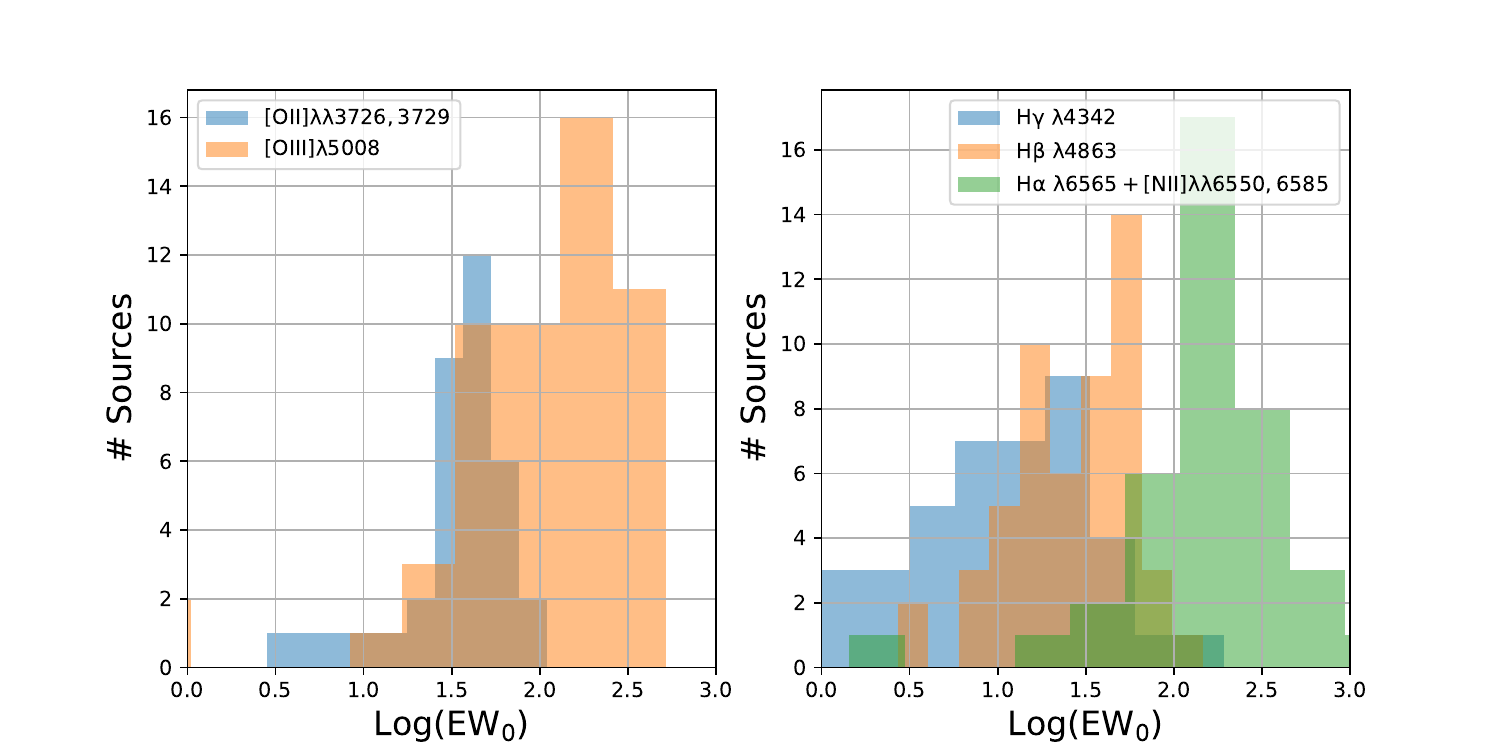}
\caption{The rest-frame equivalent widths (EW) of the \OIIg\ and \OIII5008  emission lines (Left panel) and of the \Hag, \Hb, and \Hg\ emission lines  (Right panel) in \NGDEEPA . \label{EW}}
\end{figure*}

Table \ref{tab:limits} lists the calculated $3\sigma$\ and $5\sigma$\ flux limits based on the data shown in  Figure \ref{SNR}. On average, \NGDEEPA\ reaches down to $1.8\ (3.3) \times 10^{-18}\ erg/s/cm^2$\ with an SNR value of 3 (5).
Figure \ref{SNR} also shows the predicted signal to noise (SNR) from the 2D simulations (Created as part of the NGDEEP proposal design), as well as from the JWST Exposure Time Calculator (ETC) 2.0 (Adjusted to account for the data having half the depth of our final goal). Overall, \NGDEEPA\ matches the pre-launch expectations well, although it is slightly less sensitive in the F115W filter than in the F150W and F200W filters. Table \ref{tab:slines} lists the number of emissions lines (SNR$>2$) identified so far in the \NGDEEPA\ data as part of this initial sample while Table \ref{all_lines} lists the individual measured emission line fluxes for the current sample. Figure \ref{EW} shows the distribution of rest-frame equivalent width for the oxygen and hydrogen lines.

\begin{deluxetable}{ccc} 
\tablewidth{0pt} 
\tablecaption{\NGDEEPA\ Emission Line Flux Limits \label{tab:limits}} 
\tablehead{ \colhead{Filter} & \colhead{$5 \sigma$\ flux limit} & \colhead{$3 \sigma$\ flux limit} \\
 & ${\rm 10^{-18}\ erg/s/cm^2}$ & ${\rm 10^{-18}\ erg/s/cm^2}$}
\startdata 
F115W & 4.17 & 2.17 \\
F150W & 2.69 & 1.49 \\
F200W & 2.83 & 1.56 \\
ALL & 3.33 & 1.84 \\
\enddata 
\end{deluxetable}

\begin{deluxetable}{ccc} 
\tablewidth{0pt} 
\tablecaption{\NGDEEPA\ Emission Lines \label{tab:slines}} 
\tablehead{ \colhead{Line} & \colhead{Number} & \colhead{Average Line Flux} \\
Identified &of Galaxies  & ${\rm 10^{-18}\ erg/s/cm^2}$}
\startdata 
${\rm [NeIV] \lambda}$2439 & 3 & 6.16 $\pm$ 1.80  \\
${\rm MgII \lambda}$2799 & 2 & 15.87 $\pm$ 5.05  \\
${\rm [NeV] \lambda\lambda}$3347 & 10 & 2.78 $\pm$ 0.26  \\
${\rm [NeV] \lambda\lambda}$3427 & 10 & 2.40 $\pm$ 5.65  \\
\OIIg & 80 & 7.13 $\pm$ 0.13  \\
${\rm [NeIII] \lambda\lambda}$3869 & 46 & 3.03 $\pm$ 0.11  \\
${\rm [NeIII] \lambda\lambda}$3890 & 28 & 1.86 $\pm$ 0.07  \\
${\rm [NeIII] \lambda\lambda}$3967 & 40 & 2.51 $\pm$ 0.30  \\
${\rm H\gamma \lambda\lambda}$4342  & 48 & 3.26 $\pm$ 0.74  \\
\Hb & 122 & 5.24 $\pm$ 0.09  \\
${\rm [OIII] \lambda\lambda}$4960 & 113 & 7.84 $\pm$ 0.11  \\
${\rm [OIII] \lambda}$5008 & 123 & 21.65 $\pm$ 0.24  \\
${\rm HeI \lambda}$5876 & 42 & 2.60 $\pm$ 1.04  \\
${\rm [OI] \lambda\lambda}$6302 & 28 & 2.64 $\pm$ 0.52  \\
\Hag & 104 & 17.14 $\pm$ 0.28  \\
${\rm [SII] \lambda\lambda}$6718 & 42 & 6.91 $\pm$ 0.23  \\
${\rm [SII] \lambda\lambda}$6732 & 12 & 4.71 $\pm$ 1.01  \\
${\rm [ArIII] \lambda\lambda}$7753 & 15 & 2.26 $\pm$ 2.81  \\
${\rm [SIII] \lambda\lambda}$9069 & 25 & 3.35 $\pm$ 0.22  \\
${\rm [SIII] \lambda\lambda}$9531 & 23 & 7.21 $\pm$ 0.93  \\
${\rm HeI \lambda}$10830 & 4 & 5.54 $\pm$ 1.67  \\
${\rm P\gamma \lambda}$12821 & 2 & 4.80 $\pm$ 1.39  \\
\enddata 
\tablecomments{Number of individual emission lines (SNR$>2$) measured in \NGDEEPA. For each emission line, we also list the mean and standard deviation of the sample of measured fluxes.}
\end{deluxetable}

\begin{deluxetable}{lllll} \label{alllines}
\tabletypesize{\tiny}\tablecaption{\NGDEEP\ Measured emission line fluxes.}\tablehead{  \colhead{ID} & \colhead{Line} & \colhead{Reshift}  & \colhead{Flux} & \colhead{${\rm EW_0}$}\\ 
 \colhead{  } & \colhead{ } &  \colhead{ } & \colhead{$10^{-19}erg/s/cm^2$} &  \colhead{$\AA$}} 
\startdata 
\multicolumn{5}{c}{AGN Candidates}\\ 
\hline 
27914 & \OIIg & 2.08 & $199.36 _{-14.48}^{+13.36}$ & $-91.11 _{-5.33}^{+5.77}$\\ 
27914 & ${\rm [NeIII]\ \lambda\lambda}$3869 & 2.08 & $70.88 _{-20.81}^{+17.35}$ & $-33.25 _{-7.94}^{+8.55}$\\ 
27914 & ${\rm [NeIII]\  \lambda\lambda}$3967 & 2.08 & $67.00 _{-11.83}^{+12.00}$ & $-31.47 _{-4.89}^{+4.99}$\\ 
27914 & \Hb & 2.08 & $66.43 _{-6.66}^{+6.49}$ & $-30.96 _{-2.51}^{+2.73}$\\ 
27914 & ${\rm [OIII]\  \lambda\lambda}$4959 & 2.08 & $39.00 _{-7.20}^{+6.55}$ & $-18.25 _{-2.79}^{+2.79}$\\ 
27914 & ${\rm [OIII]\  \lambda}$5006 & 2.08 & $178.28 _{-9.00}^{+9.53}$ & $-83.32 _{-3.99}^{+3.64}$\\ 
27914 & ${\rm HeI\  \lambda}$5876 & 2.08 & $46.40 _{-7.77}^{+8.16}$ & $-22.37 _{-3.21}^{+3.28}$\\ 
27914 & \Hag & 2.08 & $720.40 _{-10.99}^{+11.10}$ & $-367.69 _{-5.40}^{+5.39}$\\ 
27914 & ${\rm [SII]\  \lambda\lambda}$6716 & 2.08 & $48.84 _{-17.52}^{+17.67}$ & $-26.11 _{-7.80}^{+7.93}$\\ 
27914 & ${\rm [SII]\  \lambda\lambda}$6730 & 2.08 & $38.95 _{-16.85}^{+18.14}$ & $-19.95 _{-8.40}^{+7.95}$\\ 
32038 & ${\rm MgII\ \lambda}$2799 & 3.18 & $244.78 _{-9.03}^{+8.55}$ & $-147.16 _{-4.88}^{+4.59}$\\ 
32038 & 3346 & 3.18 & $9.73 _{-3.44}^{+3.28}$ & $-7.11 _{-2.03}^{+2.19}$\\ 
32038 & ${\rm [NeV]\ \lambda\lambda}$3427 & 3.18 & $25.99 _{-3.16}^{+3.27}$ & $-19.76 _{-2.12}^{+2.01}$\\ 
32038 & \OIIg & 3.18 & $14.55 _{-3.40}^{+3.47}$ & $-11.85 _{-2.53}^{+2.49}$\\ 
32038 & ${\rm [NeIII]\ \lambda\lambda}$3869 & 3.18 & $36.18 _{-3.70}^{+3.85}$ & $-30.21 _{-2.64}^{+2.79}$\\ 
32038 & ${\rm [NeIII]\  \lambda\lambda}$3890 & 3.18 & $12.83 _{-3.49}^{+3.44}$ & $-10.97 _{-2.42}^{+2.52}$\\ 
32038 & ${\rm [NeIII]\  \lambda\lambda}$3967 & 3.18 & $19.85 _{-4.47}^{+4.26}$ & $-17.88 _{-3.67}^{+3.52}$\\ 
32038 & ${\rm H\gamma\  \lambda\lambda}$4342 & 3.18 & $52.16 _{-4.73}^{+4.71}$ & $-42.71 _{-3.38}^{+3.48}$\\ 
32038 & \Hb & 3.18 & $109.10 _{-5.87}^{+5.62}$ & $-92.64 _{-4.15}^{+4.52}$\\ 
32038 & ${\rm [OIII]\  \lambda\lambda}$4959 & 3.18 & $121.77 _{-5.34}^{+5.25}$ & $-105.49 _{-3.76}^{+4.22}$\\ 
32038 & ${\rm [OIII]\  \lambda}$5006 & 3.18 & $306.02 _{-6.06}^{+6.11}$ & $-265.95 _{-4.67}^{+4.66}$\\ 
\hline 
\multicolumn{5}{c}{Star Forming Galaxies}\\ 
\hline 
33273 & \OIIg & 1.85 & $31.22 _{-7.50}^{+7.65}$ & $-123.71 _{-26.37}^{+26.70}$\\ 
33273 & ${\rm [NeIII]\ \lambda\lambda}$3869 & 1.85 & $22.45 _{-8.59}^{+8.17}$ & $-83.35 _{-26.01}^{+26.13}$\\ 
33273 & ${\rm [NeIII]\  \lambda\lambda}$3890 & 1.85 & $54.58 _{-8.04}^{+8.09}$ & $-200.45 _{-25.75}^{+26.35}$\\ 
33273 & ${\rm [NeIII]\  \lambda\lambda}$3967 & 1.85 & $23.60 _{-5.79}^{+6.06}$ & $-80.53 _{-18.74}^{+16.36}$\\ 
33273 & ${\rm H\gamma\  \lambda\lambda}$4342 & 1.85 & $62.35 _{-5.14}^{+5.30}$ & $-189.75 _{-14.30}^{+14.09}$\\ 
33273 & \Hb & 1.85 & $109.43 _{-3.36}^{+3.41}$ & $-266.57 _{-7.61}^{+6.80}$\\ 
33273 & ${\rm [OIII]\  \lambda\lambda}$4959 & 1.85 & $254.58 _{-3.59}^{+3.61}$ & $-598.11 _{-8.48}^{+8.51}$\\ 
33273 & ${\rm [OIII]\  \lambda}$5006 & 1.85 & $635.79 _{-4.45}^{+4.34}$ & $-1477.70 _{-10.89}^{+10.66}$\\ 
33273 & \Hag & 1.85 & $313.89 _{-5.83}^{+5.82}$ & $-1311.11 _{-21.10}^{+21.23}$\\ 
35875 & 3346 & 2.23 & $56.43 _{-18.81}^{+18.64}$ & $-13.68 _{-4.13}^{+4.00}$\\ 
35875 & \OIIg & 2.23 & $633.01 _{-16.24}^{+16.36}$ & $-160.68 _{-4.00}^{+4.13}$\\ 
35875 & ${\rm [NeIII]\ \lambda\lambda}$3869 & 2.23 & $294.06 _{-15.28}^{+14.84}$ & $-78.89 _{-3.74}^{+3.73}$\\ 
35875 & ${\rm H\gamma\  \lambda\lambda}$4342 & 2.23 & $196.45 _{-6.60}^{+7.06}$ & $-49.23 _{-1.60}^{+1.60}$\\ 
35875 & \Hb & 2.23 & $392.48 _{-7.76}^{+7.57}$ & $-104.20 _{-2.17}^{+2.12}$\\ 
35875 & ${\rm [OIII]\  \lambda\lambda}$4959 & 2.23 & $842.90 _{-7.91}^{+8.00}$ & $-234.36 _{-2.99}^{+2.85}$\\ 
35875 & ${\rm [OIII]\  \lambda}$5006 & 2.23 & $1752.73 _{-9.49}^{+9.77}$ & $-500.23 _{-7.83}^{+9.37}$\\ 
35875 & ${\rm HeI\  \lambda}$5876 & 2.23 & $55.04 _{-9.09}^{+9.12}$ & $-20.95 _{-2.90}^{+2.98}$\\ 
35875 & ${\rm [OI]\  \lambda\lambda}$6302 & 2.23 & $33.35 _{-11.11}^{+11.06}$ & $-15.63 _{-4.39}^{+4.66}$\\ 
35875 & \Hag & 2.23 & $1514.00 _{-15.19}^{+14.74}$ & $-833.88 _{-10.85}^{+10.73}$\\ 
35875 & ${\rm [SII]\  \lambda\lambda}$6716 & 2.23 & $206.67 _{-11.52}^{+11.51}$ & $-127.06 _{-6.25}^{+6.54}$\\ 
26465 & \Hb & 1.30 & $107.06 _{-8.30}^{+8.45}$ & $-147.48 _{-10.48}^{+10.48}$\\ 
26465 & ${\rm [OIII]\  \lambda}$5006 & 1.30 & $351.54 _{-12.45}^{+12.17}$ & $-502.31 _{-16.35}^{+17.18}$\\ 
26465 & ${\rm HeI\  \lambda}$5876 & 1.30 & $13.12 _{-5.39}^{+5.27}$ & $-26.18 _{-9.66}^{+9.10}$\\ 
26465 & ${\rm [OI]\  \lambda\lambda}$6302 & 1.30 & $23.37 _{-4.26}^{+3.94}$ & $-49.87 _{-7.47}^{+7.96}$\\ 
26465 & \Hag & 1.30 & $191.90 _{-5.01}^{+5.11}$ & $-457.72 _{-10.69}^{+10.32}$\\ 
26465 & ${\rm [SIII]\  \lambda\lambda}$9069 & 1.30 & $18.86 _{-6.41}^{+7.12}$ & $-86.49 _{-28.63}^{+26.92}$\\ 
26465 & ${\rm [SIII]\  \lambda\lambda}$9531 & 1.30 & $21.38 _{-7.14}^{+8.52}$ & $-116.48 _{-36.16}^{+37.29}$\\ 
27581 & ${\rm [NeV]\ \lambda\lambda}$3427 & 2.68 & $22.31 _{-8.44}^{+8.85}$ & $-36.69 _{-12.13}^{+12.53}$\\ 
27581 & \OIIg & 2.68 & $69.83 _{-5.29}^{+5.25}$ & $-87.23 _{-5.98}^{+5.66}$\\ 
27581 & ${\rm [NeIII]\ \lambda\lambda}$3869 & 2.68 & $13.59 _{-4.15}^{+4.25}$ & $-15.78 _{-4.04}^{+4.44}$\\ 
27581 & ${\rm H\gamma\  \lambda\lambda}$4342 & 2.68 & $12.67 _{-4.16}^{+4.23}$ & $-15.11 _{-4.09}^{+4.28}$\\ 
27581 & \Hb & 2.68 & $38.70 _{-7.00}^{+7.53}$ & $-52.52 _{-9.13}^{+7.99}$\\ 
27581 & ${\rm [OIII]\  \lambda\lambda}$4959 & 2.68 & $48.99 _{-6.21}^{+6.37}$ & $-66.62 _{-8.07}^{+7.66}$\\ 
27581 & ${\rm [OIII]\  \lambda}$5006 & 2.68 & $114.52 _{-7.03}^{+7.02}$ & $-158.51 _{-8.58}^{+8.62}$\\ 
27581 & ${\rm HeI\  \lambda}$5876 & 2.68 & $17.14 _{-5.38}^{+5.58}$ & $-36.29 _{-10.60}^{+10.23}$\\ 
\enddata  
\end{deluxetable}

\subsection{Spectral Energy Distribution Analysis}\label{sec:SED}

\begin{figure*}
\center
\includegraphics[width=6.5in]{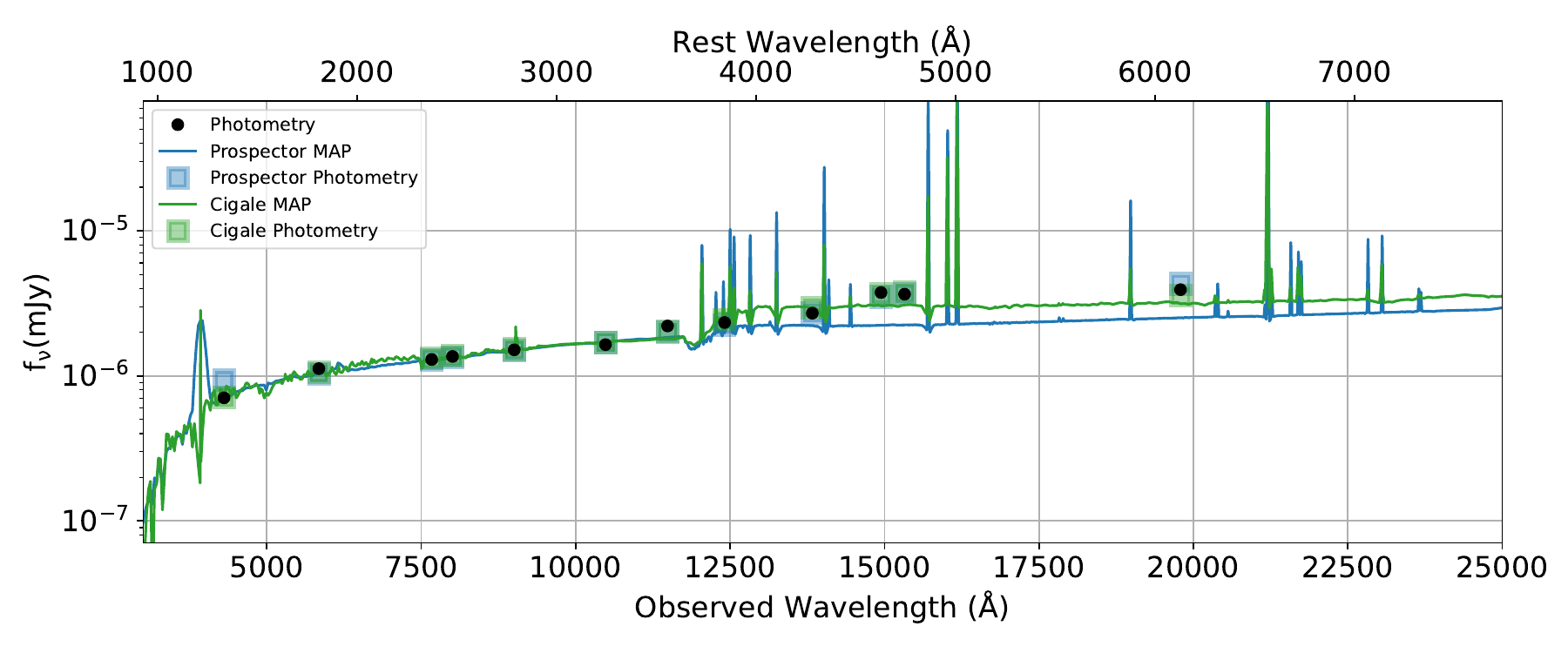}

\caption{Prospector and CIGALE maximum a posteriori  fits for the source (ID$\#$35875) shown in Figure \ref{2Dshift}.  The results from Prospector produces a} stellar mass of this object $\approx 10^{10}\ M_\sun$ with a stellar age of $\approx 0.1$\ Gyr old .\label{SED}
\end{figure*}

\begin{figure*}
\center
\includegraphics[width=7.5in]{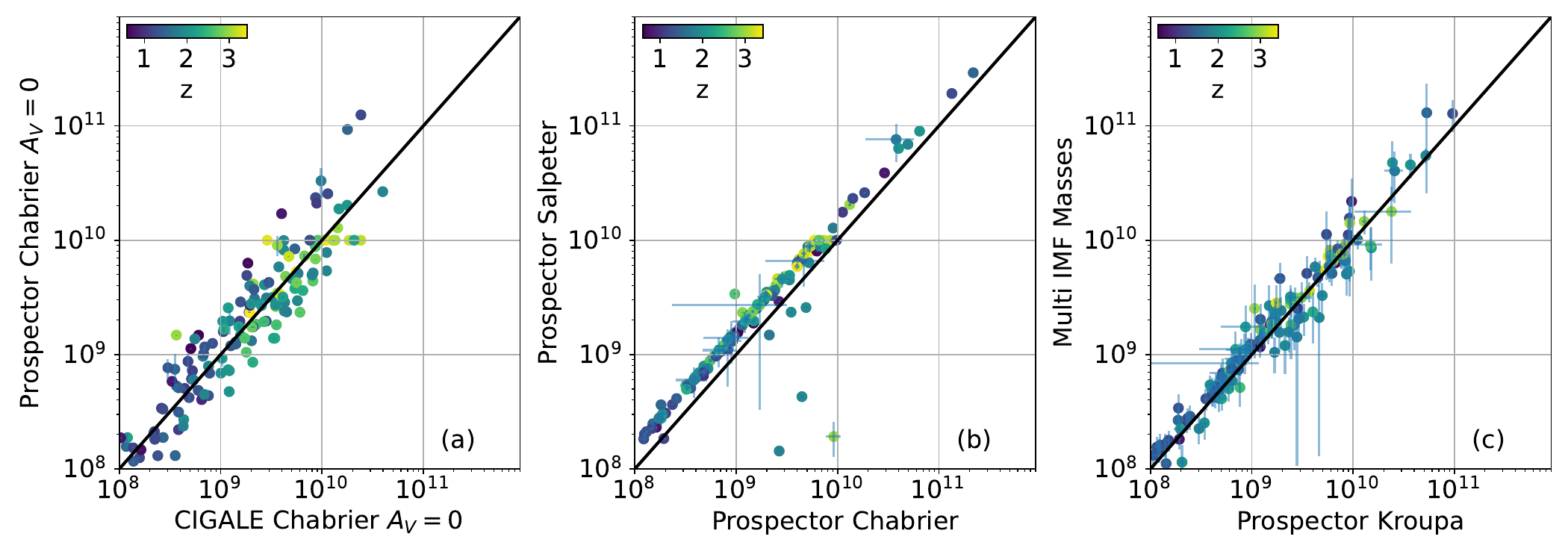}
\caption{ Comparison of the stellar masses estimates derived using CIGALE and Prospector.  In each panel, the diagonal line represents a 1:1 match in the data values.  Panel (a) compares the masses derived assuming no dust and a Chabrier IMF. The scatter of the points, errors, and distance relative to the 1:1 line are similar when a Salpeter IMF (without dust) is used to compute stellar masses for CIGALE and Prospector.  Panel (b) shows a comparison between the Salpeter and Chabrier IMFs using only Prospector (again, without dust).  This illustrates the impact to SED derived stellar masses. Panel (c) shows results from Prospector alone, and compares the errors associated from computing SEDs using three different IMFs, and other parameters discussed in the text (Y-axis), against the errors generated from assuming a single IMF (Kroupa) and power law dust law (essentially the default parameters for computing an SED in Prospector).  The panel demonstrates a comparison between a more realistic approach to errors and stellar mass values (Y-axis) and simply choosing one set of values and using Prospector's internal errors (X-axis).  \label{SED_ALL}}
\end{figure*}

\begin{figure*}
\center
\includegraphics[width=7.5in]{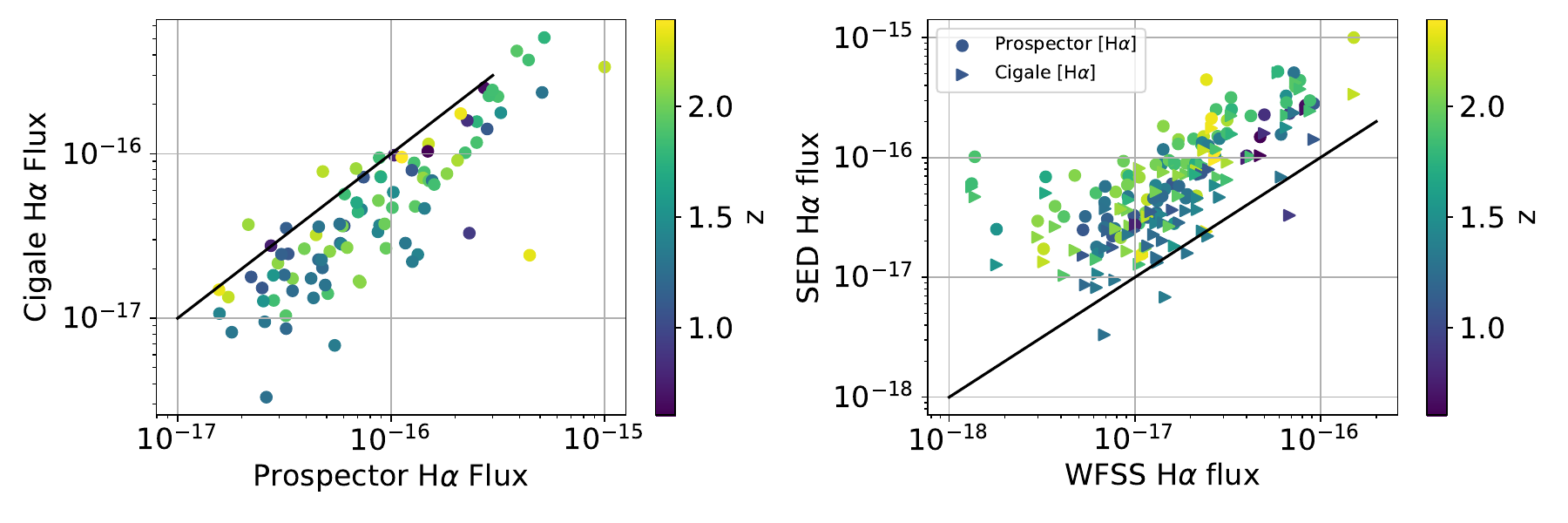}
\caption{ Comparison of the \Hag\ emission line fluxes used as part of the nebular emission models used by CIGALE, Prospector, and actual \Hag\ line fluxes measured by the \NGDEEPA\ survey. Left Panel: CIGALE \Hag\ emission line fluxes versus Prospector \Hag\ emission line fluxes, color coded by redshift, both using a Chabrier IMF. Prospector usually predicts higher line fluxes than CIGALE. Right Panel: The Prospector (circle) and CIGALE (triangle) emission line fluxes, using a Chabrier IMF, plotted versus the measured \NGDEEPA\ \Hag\ fluxes. The predicted emission line fluxes are systematically above those actually measured from the real data. \label{SEDfluxplot}}
\end{figure*}

 In addition to using the \NGDEEPA\ to determine physical parameters of the sample using emission lines, the photometric data from HST (ACS and WFC3-IR) and JWST/NRISS observations across 12 filters were used to construct SEDs. Table \ref{tab:phot} provides the photometric parameters for the \NGDEEP\ galaxies.  The resulting SEDs were then used primarily to estimate stellar mass and internal extinction from dust, but also provided some estimates of star-formation histories (ages and rates). The outputted parameters from such approaches are strongly dependent upon several factors (in order of impact to the final result):  how the model parameter space is sampled, optimized, and fit to the observational data; the assumed initial mass function (IMF); and the application of particular dust laws (and the assumptions associated with their use, such as recombination cases).  Taken together, these produce a scatter in the resulting stellar masses of a few tens of percent. We started by using the Code Investigating GALaxy Emission \citep[Cigale,][]{Burgarella05,Noll09,Boquien19}, which has the advantage of being fast but the drawback of depending on a fixed pre-defined grid of model parameters and therefore not able to produce full posterior distributions for the fitted model parameters. Figure \ref{SED} shows an example of the photometric SEDs generated from both Prospector (Using a Kroupa IMF, a delayed star formation history, and power law dust model with an index of -0.7) and CIGALE (Using a Chabrier IMF, a delayed star formation history, and a \citet{Charlot00}, CF00, extinction law) for one source in the sample ($m_{AB, F160W}=22.48$ galaxy at z=2.23).  The figure shows both the HST and JWST photometry plotted along with the best matched SEDs, and the photometry in the same filters as estimated from the SEDs and input parameters.  Figure \ref{SED} also shows the maximum {\it a posteriori}  (MAP) models for the spectra shown in Figure \ref{2Dshift}. 

 Prospector \citep{Prospector21} was selected because it provides a convenient way to combine complex stellar population models \citep[Flexible Stellar Population Synthesis, FSPS, ][]{Conroy09,Conroy10} with a forward modelling method, albeit at the expense of much longer running time. However, we first verified that the stellar masses derived using both codes were consistent with each other using two different IMFs (Salpeter and Chabrier) and without applying any dust laws.  This was done because the two codes do not provide the same choices for dust laws, and adding these differences would have introduced additional systematics. Panel (a) of Figure \ref{SED_ALL} shows the comparison between Prospector and CIGALE for the same IMF and without any dust laws applied.  The same test was run for the Salpeter IMF and produced similar scatter along the 1:1 correlation line plotted in the panel.  Both codes produce error bars which are significantly smaller than the scatter in the data between them.  It was also found that Prospector and CIGALE produced best fit models to our observations in which the nebular emission line fluxes differed significantly from each other.  CIGALE emission line fluxes were systematically smaller than those generated with Prospector (Figure \ref{SEDfluxplot} left panel).  More concerning, however, is that {\it both} codes were {\it systematically and significantly different from the actual WFSS measured line flux measurements}.  Both CIGALE and Prospector produced SED generated emission line fluxes significantly brighter than those measured from the actual data. This is shown in the right panel of Figure \ref{SEDfluxplot}. Both codes significantly over-predict the \Hag\ fluxes by as much as 0.5 dex. Furthermore, there is an indication that the emission line fluxes are systematically over-predicted as redshift increases. 

As noted earlier, the choice of the IMF has an impact on the stellar masses derived from fits to the SEDs. 
Panel (b) of Figure \ref{SED_ALL} demonstrates how the errors, even when derived from the full posterior distribution do not reflect the actual uncertainties in stellar mass resulting from different IMF assumptions. For the purpose of this paper, and in order to derive stellar masses with realistic errors based on legitimate uncertainties from having no direct knowledge of the physical processes occurring in these galaxies, Prospector SEDs were generated multiple times.  We used Salpeter, Chabrier, and Kroupa IMFs, using both a power law dust law with an index of -0.7 as well as using the \citet{Calzettti00} dust law (C00). Flat posteriors were used for all model parameters. Dust content, metallicity, and mass were all allowed to vary and a delayed exponential star formation history was used assumed (${\rm SFR(t) \propto t \times e^{-t/\tau}}$).  Computed mean masses and errors were generated based on these multiple, independent analyses. The final stellar masses for the \NGDEEPA\ sample and associated error bars is shown in Panel (c) of Figure \ref{SED_ALL}.
Table \ref{tab:propsector} lists some of the average physical parameters for each of the \NGDEEP\ galaxies, as determined using Prospector. 
\newpage
\startlongtable
\movetabledown=2in
\begin{rotatetable}
\begin{deluxetable}{lllllllllllllll}
\tabletypesize{\tiny}\tablecaption{\NGDEEP\ Emission line galaxy photometric measurements.}\label{tab:phot} 
\tablehead{  \colhead{ID} & \colhead{RA} & \colhead{DEC} & \colhead{F435W} & \colhead{F606W} & \colhead{F775W} & \colhead{F814W} & \colhead{F850LP} & \colhead{F105W} & \colhead{F125W} & \colhead{F140W} & \colhead{F160W} & \colhead{F115W} & \colhead{F150W} & \colhead{F200W}\\ 
\colhead{ } & \colhead{deg} & \colhead{deg} & \colhead{nJy} & \colhead{nJy} & \colhead{nJy} & \colhead{nJy} & \colhead{nJy} & \colhead{nJy} & \colhead{nJy} & \colhead{nJy} & \colhead{nJy} & \colhead{nJy} & \colhead{nJy} & \colhead{nJy}} 
\startdata 
\multicolumn{15}{c}{AGN Candidates}\\ 
\hline 
 27914  & 53.169713  & -27.797083  & $153.5\pm0.7$  & $262.7\pm0.4$  & $362.0\pm0.3$  & $414.5\pm4.6$  & $519.0\pm0.6$  & $767.1\pm0.9$  & $1378.3\pm0.7$  & $1739.8\pm0.2$  & $2000.3\pm0.6$  & $1121.2\pm13.8$  & $2012.4\pm24.9$  & $3389.8\pm33.8$ \\ 
 32038  & 53.178500  & -27.784102  & $188.7\pm0.5$  & $647.8\pm0.3$  & $585.7\pm0.2$  & $821.8\pm4.3$  & $641.0\pm0.5$  & $998.4\pm0.8$  & $1194.3\pm0.6$  & $1472.8\pm0.1$  & $1386.7\pm0.5$  & $964.3\pm11.3$  & $1158.3\pm18.6$  & $1816.3\pm23.3$ \\ 
\hline 
\multicolumn{15}{c}{Star Forming Galaxies}\\ 
\hline 
 33273  & 53.153440  & -27.781108  & $161.5\pm0.1$  & $160.0\pm0.1$  & $148.5\pm0.0$  & $154.5\pm0.8$  & $155.8\pm0.1$  & $168.8\pm0.1$  & $176.8\pm0.1$  & $339.8\pm0.0$  & $400.9\pm0.1$  & $176.7\pm5.9$  & $434.6\pm10.7$  & $241.8\pm9.2$ \\ 
 35875  & 53.154482  & -27.771513  & $695.5\pm0.3$  & $1106.5\pm0.2$  & $1278.7\pm0.1$  & $1341.7\pm2.7$  & $1491.0\pm0.3$  & $1618.5\pm0.4$  & $2300.6\pm0.3$  & $2665.4\pm0.1$  & $3605.8\pm0.3$  & $2176.5\pm15.5$  & $3701.5\pm30.8$  & $3873.4\pm29.8$ \\ 
 26465  & 53.147728  & -27.804064  & $169.6\pm1.5$  & $175.6\pm1.0$  & $199.7\pm1.0$  & $230.2\pm4.3$  & $315.9\pm1.9$  & $392.0\pm5.7$  & $411.5\pm6.0$  & $419.1\pm24.2$  & $428.0\pm7.6$  & $414.8\pm6.3$  & $491.6\pm9.7$  & $466.2\pm9.4$ \\ 
 27581  & 53.156369  & -27.799041  & $182.8\pm0.5$  & $331.0\pm0.3$  & $367.3\pm0.2$  & $370.0\pm4.2$  & $391.2\pm0.5$  & $406.1\pm0.7$  & $494.3\pm0.6$  & $650.3\pm0.1$  & $822.2\pm0.5$  & $444.3\pm9.6$  & $863.4\pm17.6$  & $1132.7\pm19.8$ \\ 
 29659  & 53.153112  & -27.792459  & $35.2\pm0.3$  & $97.1\pm0.2$  & $127.2\pm0.1$  & $129.3\pm2.1$  & $122.1\pm0.2$  & $121.3\pm0.4$  & $126.1\pm0.3$  & $127.9\pm0.1$  & $153.4\pm0.2$  & $125.0\pm7.4$  & $156.0\pm11.9$  & $292.9\pm12.6$ \\ 
 33017  & 53.167466  & -27.781899  & $90.5\pm0.4$  & $101.7\pm0.2$  & $98.5\pm0.2$  & $103.8\pm2.9$  & $99.5\pm0.3$  & $117.6\pm0.6$  & $156.7\pm0.4$  & $236.5\pm0.1$  & $281.6\pm0.3$  & $142.9\pm5.2$  & $286.9\pm9.8$  & $216.6\pm8.9$ \\ 
 34323  & 53.146655  & -27.777558  & $209.2\pm0.9$  & $221.3\pm0.6$  & $282.1\pm0.4$  & $310.1\pm6.3$  & $348.2\pm1.0$  & $386.3\pm1.2$  & $370.8\pm0.9$  & $382.6\pm0.2$  & $357.8\pm0.8$  & $407.6\pm9.6$  & $443.8\pm14.5$  & $440.0\pm15.1$ \\ 
 34389  & 53.144302  & -27.777234  & $180.2\pm1.1$  & $193.3\pm0.7$  & $282.7\pm0.5$  & $323.4\pm7.6$  & $438.9\pm1.2$  & $456.3\pm1.7$  & $472.4\pm1.4$  & $477.3\pm0.3$  & $486.0\pm1.2$  & $513.2\pm9.7$  & $566.9\pm15.0$  & $587.6\pm14.9$ \\ 
 35250  & 53.159595  & -27.774640  & $109.6\pm1.0$  & $139.1\pm0.6$  & $182.3\pm0.4$  & $203.2\pm7.7$  & $220.0\pm1.0$  & $348.7\pm1.5$  & $534.2\pm1.1$  & $554.5\pm0.3$  & $544.9\pm1.0$  & $483.4\pm9.3$  & $647.4\pm15.2$  & $779.9\pm16.2$ \\ 
 36241  & 53.162851  & -27.771691  & $72.1\pm1.1$  & $202.9\pm0.6$  & $288.6\pm0.5$  & $284.0\pm8.8$  & $316.8\pm1.1$  & $301.7\pm1.5$  & $320.8\pm1.2$  & $321.9\pm0.3$  & $399.9\pm1.1$  & $320.9\pm5.5$  & $407.0\pm9.0$  & $814.0\pm10.5$ \\ 
 37266  & 53.165945  & -27.767997  & $42.0\pm0.8$  & $121.7\pm0.5$  & $168.8\pm0.3$  & $159.4\pm6.1$  & $179.1\pm0.8$  & $171.3\pm1.0$  & $171.6\pm0.8$  & $178.3\pm0.2$  & $189.3\pm0.7$  & $178.1\pm8.2$  & $227.2\pm13.0$  & $422.2\pm15.1$ \\ 
 38883  & 53.167361  & -27.762843  & $71.6\pm0.4$  & $96.2\pm0.3$  & $145.7\pm0.2$  & $164.5\pm3.5$  & $201.7\pm0.4$  & $172.9\pm1.1$  & $194.1\pm0.9$  & $175.5\pm0.5$  & $174.3\pm0.8$  & $240.1\pm4.0$  & $238.8\pm6.3$  & $216.7\pm8.9$ \\
\enddata
\tablecomments{The two objects detected in the X-ray \citep{Luo17} with fluxes $\ge$ 10$^{42}$ erg s$^{-1}$ are listed at the top part of this Table. Full table is available online}
\end{deluxetable} 

\end{rotatetable}
\newpage

\begin{deluxetable}{cccccccc}
\tabletypesize{\scriptsize}\tablecaption{Prospector derived parameters.}\label{tab:propsector} 
\tablehead{  \colhead{ID} & \colhead{z} & \colhead{Log(Mass)} & \colhead{Z} & \colhead{Av} & \colhead{Age} & \colhead{$\tau$} & \colhead{SFR}\\ 
\colhead{ } & \colhead{ } & \colhead{Log(M$_\sun$)} & \colhead{Log(O/H)} & \colhead{mag} & \colhead{Gyr} & \colhead{Gyr} & \colhead{M$_\sun / yr$}} 
\startdata 
\multicolumn{8}{c}{AGN Candidates}\\ 
\hline 
 27914  & 2.08  & $10.66 ^{+0.09} _{-0.12}$ & $9.76 ^{+0.71} _{-0.71}$ & $1.37 ^{+0.01} _{-0.01}$ & $0.69 ^{+0.32} _{-0.31}$ & $0.26 ^{+0.21} _{-0.20}$ & $45.30 ^{+42.07} _{-41.60}$\\ 
 32038  & 3.18  & $9.55 ^{+0.08} _{-0.10}$ & $8.47 ^{+0.00} _{-0.00}$ & $1.48 ^{+0.00} _{-0.00}$ & $0.01 ^{+0.00} _{-0.00}$ & $0.10 ^{+0.00} _{-0.00}$ & $1120.87 ^{+224.40} _{-224.08}$\\ 
\hline 
\multicolumn{8}{c}{Star Forming Galaxies}\\ 
\hline 
 33273  & 1.85  & $7.45 ^{+0.10} _{-0.14}$ & $9.62 ^{+0.07} _{-0.07}$ & $0.24 ^{+0.00} _{-0.00}$ & $0.00 ^{+0.00} _{-0.00}$ & $0.11 ^{+0.00} _{-0.00}$ & $13.56 ^{+3.97} _{-3.99}$\\ 
 35875  & 2.23  & $9.95 ^{+0.17} _{-0.26}$ & $9.86 ^{+0.61} _{-0.63}$ & $1.08 ^{+0.00} _{-0.00}$ & $0.11 ^{+0.06} _{-0.06}$ & $0.10 ^{+0.00} _{-0.00}$ & $105.71 ^{+51.56} _{-50.68}$\\ 
 26465  & 1.30  & $8.93 ^{+0.10} _{-0.13}$ & $10.47 ^{+0.02} _{-0.02}$ & $0.15 ^{+0.04} _{-0.04}$ & $0.58 ^{+0.13} _{-0.13}$ & $0.22 ^{+0.08} _{-0.08}$ & $1.01 ^{+0.48} _{-0.50}$\\ 
 27581  & 2.68  & $9.87 ^{+0.09} _{-0.11}$ & $10.39 ^{+0.34} _{-0.34}$ & $0.62 ^{+0.00} _{-0.00}$ & $0.28 ^{+0.05} _{-0.05}$ & $0.10 ^{+0.00} _{-0.00}$ & $16.97 ^{+6.59} _{-6.47}$\\ 
 29659  & 3.32  & $9.45 ^{+0.16} _{-0.25}$ & $9.37 ^{+0.45} _{-0.44}$ & $0.38 ^{+0.01} _{-0.01}$ & $0.39 ^{+0.11} _{-0.11}$ & $0.10 ^{+0.00} _{-0.00}$ & $2.40 ^{+2.14} _{-2.17}$\\ 
 33017  & 2.07  & $8.03 ^{+0.13} _{-0.19}$ & $9.79 ^{+0.05} _{-0.05}$ & $0.47 ^{+0.00} _{-0.00}$ & $0.06 ^{+0.02} _{-0.02}$ & $15.70 ^{+0.90} _{-0.87}$ & $3.71 ^{+1.82} _{-1.78}$\\ 
 34323  & 1.09  & $8.41 ^{+0.08} _{-0.10}$ & $10.11 ^{+0.01} _{-0.01}$ & $0.03 ^{+0.01} _{-0.01}$ & $0.27 ^{+0.02} _{-0.02}$ & $0.11 ^{+0.01} _{-0.01}$ & $0.69 ^{+0.17} _{-0.16}$\\ 
 34389  & 1.09  & $8.94 ^{+0.08} _{-0.10}$ & $9.47 ^{+0.01} _{-0.01}$ & $0.00 ^{+0.01} _{-0.01}$ & $0.55 ^{+0.00} _{-0.00}$ & $0.10 ^{+0.00} _{-0.00}$ & $0.20 ^{+0.04} _{-0.04}$\\ 
 35250  & 1.69  & $9.77 ^{+0.10} _{-0.14}$ & $9.36 ^{+0.24} _{-0.24}$ & $0.84 ^{+0.03} _{-0.03}$ & $1.74 ^{+0.96} _{-0.97}$ & $1.82 ^{+2.44} _{-2.39}$ & $4.80 ^{+6.60} _{-6.60}$\\ 
 36241  & 3.33  & $9.86 ^{+0.10} _{-0.12}$ & $9.62 ^{+0.06} _{-0.06}$ & $0.53 ^{+0.01} _{-0.01}$ & $0.33 ^{+0.01} _{-0.01}$ & $0.10 ^{+0.00} _{-0.00}$ & $10.72 ^{+2.96} _{-2.90}$\\ 
 37266  & 3.42  & $9.77 ^{+0.09} _{-0.11}$ & $8.93 ^{+0.55} _{-0.56}$ & $0.00 ^{+0.00} _{-0.00}$ & $0.54 ^{+0.11} _{-0.10}$ & $0.10 ^{+0.00} _{-0.00}$ & $1.46 ^{+1.25} _{-1.31}$\\ 
 38883  & 0.76  & $8.26 ^{+0.07} _{-0.09}$ & $9.54 ^{+0.10} _{-0.10}$ & $0.28 ^{+0.03} _{-0.03}$ & $2.04 ^{+0.62} _{-0.62}$ & $53.68 ^{+11.59} _{-11.61}$ & $0.17 ^{+0.10} _{-0.09}$\\ 
\enddata 
\tablecomments{The two objects detected in the X-ray \citep{Luo17} with fluxes $\ge$ 10$^{42}$ erg s$^{-1}$ are listed at the top part of this Table. Full table is available online}\end{deluxetable}

\subsection{Emission line Ratios}
\subsubsection{[NII] contamination of the observed \Ha\ fluxes}\label{[NII]}
The low resolution of the NIRISS WFSS makes it impossible to separate \NIIg\ emission from nearby \Ha.  However, one can apply a statistical correction to the 104 measured \Hag\ fluxes to derive intrinsic \Ha\ fluxes. \citet{Faisst18} derived an empirical correction based on the \OIIIg, \Hb, and stellar masses:

\begin{equation}
\begin{gathered}
M(Log([NII]/H\alpha),z) = 3.696\ \xi + 3.236\ \xi^{-1} + 0.729\ \xi^{-2} + 14.928 + 0.156(1+z)^2 \\
\xi=Log([NII]/H\alpha) + 0.138 - 0.042(1+z)^2
\end{gathered}\label{eq:faisst18}
\end{equation}

\begin{figure*}
\center
\includegraphics[width=7.5in]{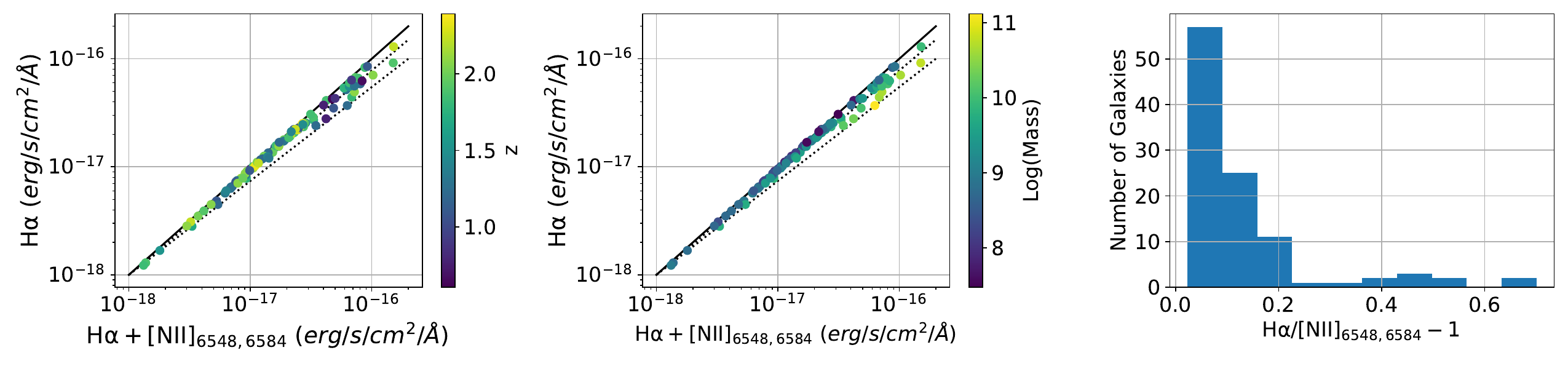}
\caption{The estimates of the \Ha\ emission line fluxes after being statistically corrected for \NIIg\ contamination following \citet{Faisst18}. We plot observed versus corrected fluxes, color coded as a function of redshift  (Left panel) and mass (Center panel). The  Right panel shows the \NIIg\ contamination as a fraction of \Ha\ flux. \label{Ha_corr}}
\end{figure*}

This relation was only calibrated up to z=2.5. While statistical in nature, it provides a more nuanced approach than simply assuming a fixed correction (e.g. 30\%) or based solely on the observed \Hag\ flux since it has now been demonstrated that the amount of \NIIg\ flux is correlated with stellar mass and redshift \citep{Faisst18}. Using this approach, it is estimated that the contribution of \NIIg\ in the sample is relatively small ($8 \pm 11$\%\ of the \Ha\ flux), as shown in Figure \ref{Ha_corr}. In the rest of this paper, all of the quoted \Ha\ fluxes have been corrected for [NII] contamination.

\subsubsection{Correcting for Extinction}\label{sec:Av}
\begin{figure*}
\center
\includegraphics[width=7in]{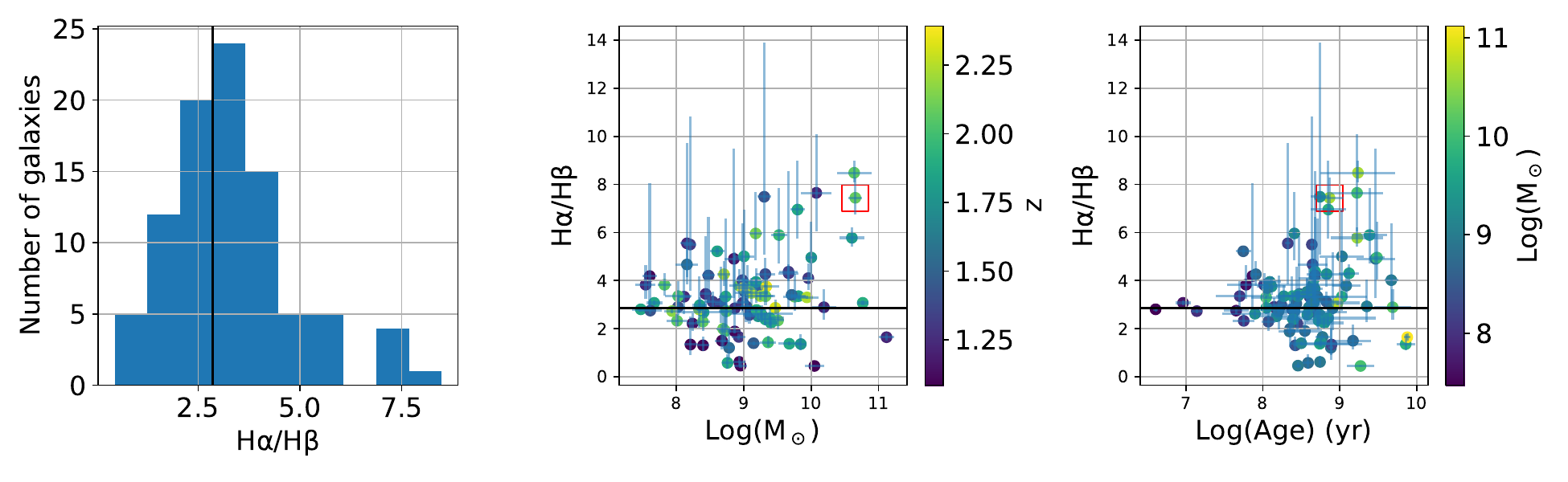}
\caption{Left Panel: Distribution of \Ha/\Hb\ lines. Center Panel:  \Ha/\Hb\ as a function of stellar mass, and color coded by redshift. Right Panel:  \Ha/\Hb\ as a function of stellar ages, and color coded by stellar mass. The canonical Case B line ratio is shown in all panels using a solid black line at a value of 2.86. X-ray AGN candidates from \citet{Luo17} are indicated using red squares.\label{HaHb}}
\end{figure*}

\begin{figure*}
\center
\includegraphics[width=7in]{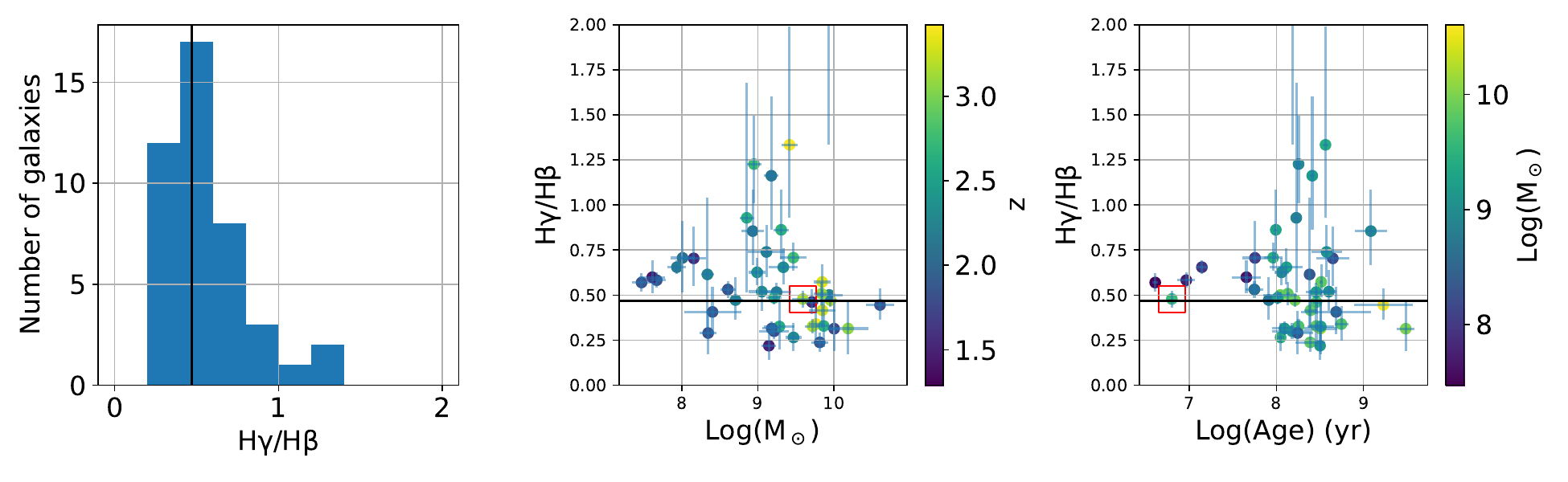}
\caption{Left Panel: Distribution of \Hg/\Hb\ lines. Center Panel:   \Hg/\Hb\ as a function of stellar mass, and color coded by redshift. Right Panel:  \Hg/\Hb\  as a function of stellar ages, and color coded by stellar mass. The canonical Case B line ratio is shown in all panels using a solid black line at a value of 0.47. X-ray AGN candidates from \citet{Luo17} are indicated using red squares. \label{HgHb}}
\end{figure*}

One can directly infer the effect of dust for a fraction of the galaxies by comparing the observed Balmer line flux ratios to their canonical values. For the purposes of this paper, Case B is used because it is generally assumed that in typical star-forming regions no Lyman $\alpha$ photons escape, and all are rescattered.  A Case B ratio of 2.86 was used for H$\alpha$/H$\beta$ and a ratio of 0.47 was used for H$\gamma$/H$\beta$ assuming a typical electron temperature of $T_e=10^4K$\ and an electron density of $n_e=10^2\ cm^{-3}$ \citep{Osterbrock89}.  For the \NGDEEPA\ sample, 73 out of 91 sources have $H\alpha$\ and  $H\beta$\ line fluxes consistent with Case B recombination, and 24 out of 45 sources have $H\gamma$\ and $H\beta$\ emission line fluxes consistent with Case B recombination.  Figures \ref{HaHb} and \ref{HgHb} show the measured values of the Balmer line ratios \Ha /\Hb\ and \Hg/\Hb.  This analysis comes with a caveat.  While the use of ``universal'' dust laws (e.g. Calzetti, LMC, SMC, etc) may be somewhat empirically successful in the local Universe, it is far from clear that blanket use of these corrections to all galaxies is warranted or justified \citep[][and references therein] {Witt2000,Salim20}.  Further, the application of such laws to more distant (e.g. z $>>$ 1) objects may not be appropriate either.  Although it is likely that many of the objects in the \NGDEEPA\ sample are dusty, the distribution of dust (uniform or clumpy), as well as the grain sizes and their composition have yet to be directly characterized or inferred, let alone demonstrated to be the same or similar to galaxies in the local Universe.  In this section, calculations of \AV\ using the Balmer decrement method are presented purely as a means to compute an average dust content in these objects and to {\it qualitatively} compare the sample to previous works as well as to illustrate the limitations of this approach.

\begin{figure*}
\center
\includegraphics[width=5in]{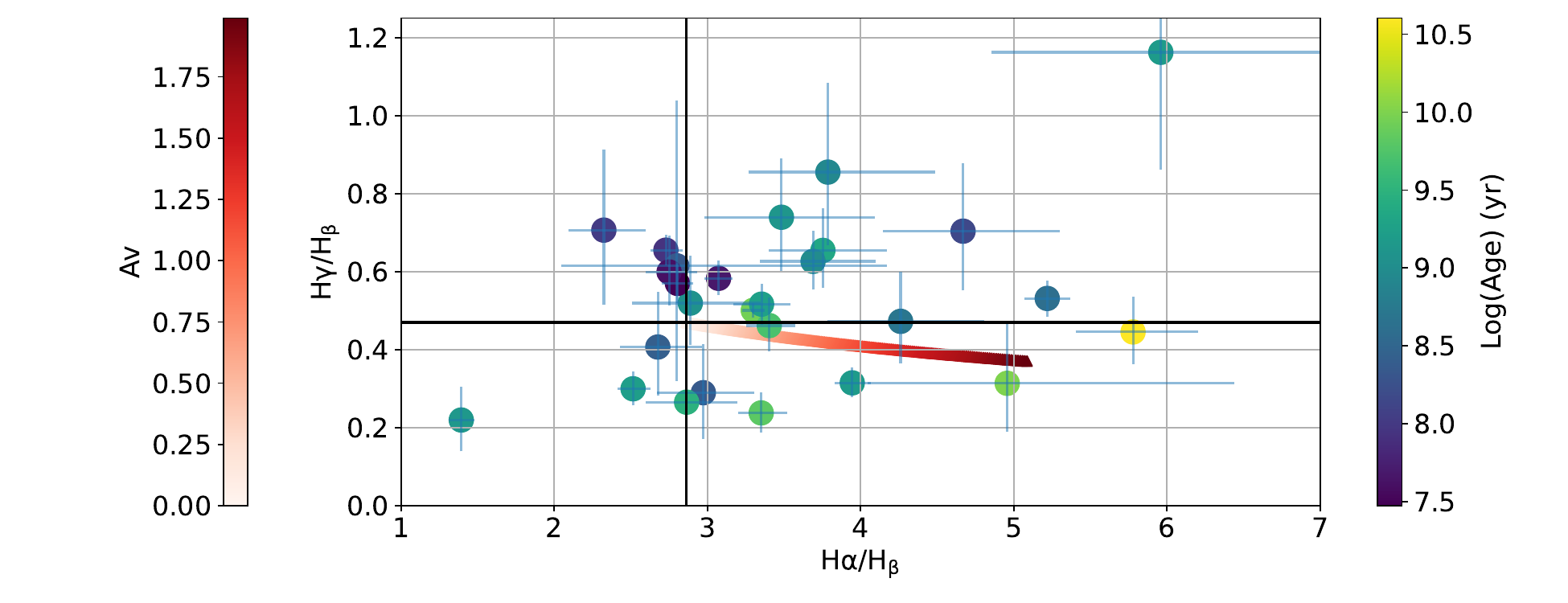}
\caption{Values of the Balmer line ratios \Ha/\Hb\ and \Hg/\Hb\ for the sample of \NGDEEPA\ galaxies with measured line fluxes for these three emission lines.  The canonical Case B line ratio is shown as a solid vertical and horizontal line for the two Balmer line ratios. We also plot line ratios for these emission lines when applying the C00 ''universal'' dust laws with values of $0<A_V<2$. Other dust laws taken from \citet{Witt2000}, which include different dust geometries, produce nearly identical line ratios.  The bottom right quadrant is where points should lie if Case B recombination were appropriate.}\label{HaHbHg}
\end{figure*}

As these figures demonstrate, 30\% of the 91 sources with a  SNR$>2$ for both \Ha\ and \Hb\ emission lines are observed to have ratios {\it lower} than the Case B value of 2.86. Moreover, 47\% of sources with \Hg\ and \Hb\ show \Hg/\Hb\ values {\it below} 0.47. The vast majority of the sample shows ratios incompatible with Case B and any ``universal'' dust law.  This result is not correlated with signal to noise or other factors related to the quality of the data. For example, Figures \ref{FITsample1} and \ref{FITsample2} clearly show detected \Hag, \Hb, and \Hg, but have \Ha/\Hb\ ratios that are either below 2.86 ($2.74^{+0.04}_{-0.11}$ and $3.15^{+0.03}_{-0.08}$, respectively) or \Hg/Hb\ that are above 0.47 ($+0.56^{0.02}_{-0.06}$\ and $0.50^{+0.01}_{-0.02}$,respectively). Figure \ref{HaHbHg} shows the distribution of Balmer line ratios for 28 sources for which there are $H\alpha$, $H\beta$. and $H\gamma$\ emission lines. As this Figure demonstrates, most of these line ratios are observed to be inconsistent with a ''universal'' dust law and a Case B photo-ionization model. 

For objects which appear to be consistent with Case B recombination, one can use the relations from \citet{Osterbrock89} to estimate the amount of extinction in the 63 galaxies using the observed \Hb\ line fluxes and the estimates of the intrinsic \Ha\ line fluxes as described in the previous Section, as well as the 24 galaxies with \Hb4861\ and $H\gamma\ \lambda4341$\ line fluxes. 
The C00 extinction curve and the equations were used \citep{Dominguez13}:

\begin{equation}
A_\lambda = \kappa(\lambda) E(B-V)
\end{equation}
with
\begin{equation}
E(B-V) = {2.5 \over {\kappa(\lambda_{H\beta}) - \kappa(\lambda_{H\alpha})}} Log[{(H\alpha/H\beta)_{obs} \over 2.86}]
\end{equation}
or
\begin{equation}
E(B-V) = {2.5 \over {\kappa(\lambda_{H\beta}) - \kappa(\lambda_{H\gamma})}} Log[{(H\gamma/H\beta)_{obs} \over 0.47}]
\end{equation}

and for the adopted C00 extinction curve ($\kappa(\lambda)$),
\begin{equation}
\kappa(V) = R_V = 4.05 \pm 0.8    \\
\end{equation}
\begin{equation}
{2.5 \over {\kappa(\lambda_{H\beta}) - \kappa(\lambda_{H\alpha})}} = 1.91 \pm 1.66   \\
\end{equation}
\begin{equation}
{2.5 \over {\kappa(\lambda_{H\beta}) - \kappa(\lambda_{H\gamma})}} = -4.76 \pm 10.3   \\
\end{equation}

\begin{figure*}
\center
\includegraphics[width=7in]{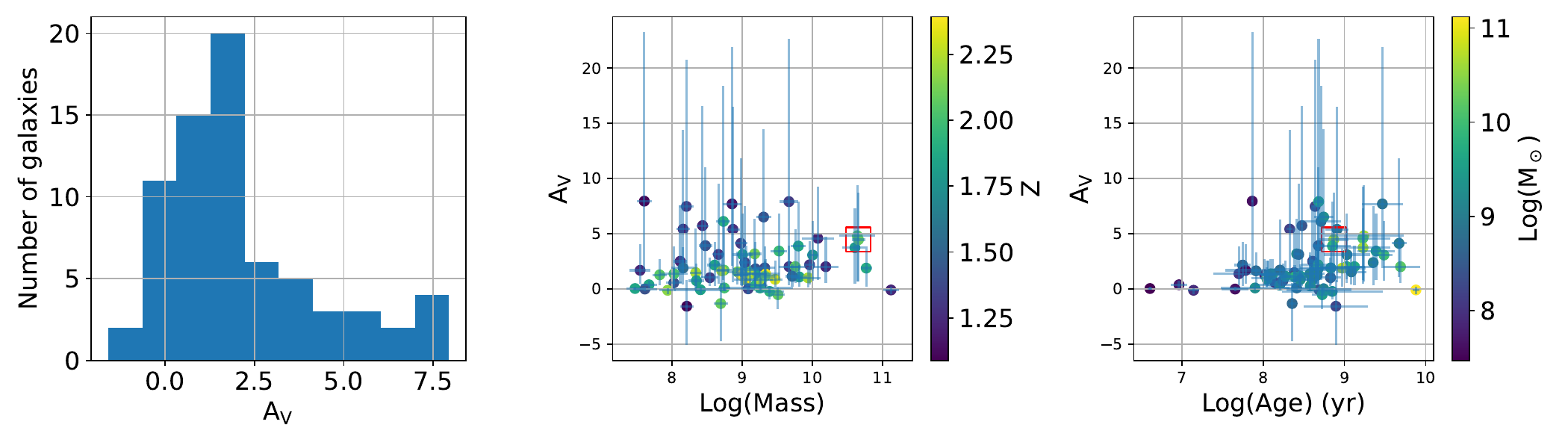}
\caption{Left Panel: Extinction (\AV) values derived using the estimated \Ha\ and measured \Hb\, assuming a Case B scenario, and an intrinsic Balmer line ratio of 2.86.  Center Panel: \AV\ as a function of stellar mass, color coded by redshift. Right Panel: \AV\ as a function of stellar ages, color coded by stellar mass.  X-ray AGN candidates from \citet{Luo17} are indicated using red squares.\label{Av}}
\end{figure*}

The left panel of Figure \ref{Av} shows the distribution of empirically derived \AV\ values using the Balmer decrement method. The center panel of this Figure shows \AV\ as a function of stellar mass (as estimated using Prospector, see Section \ref{sec:SED} for details) color coded by redshift. The right panel compares \AV\ with stellar ages (as estimated using Prospector)  and color coded by mass. The figure shows no evidence for a dependence of \AV\ on stellar mass.  However, the right panel does show a correlation between \AV\ and stellar ages, with increased \AV\ as age increases.

\begin{figure*}
\center
\includegraphics[width=4in]{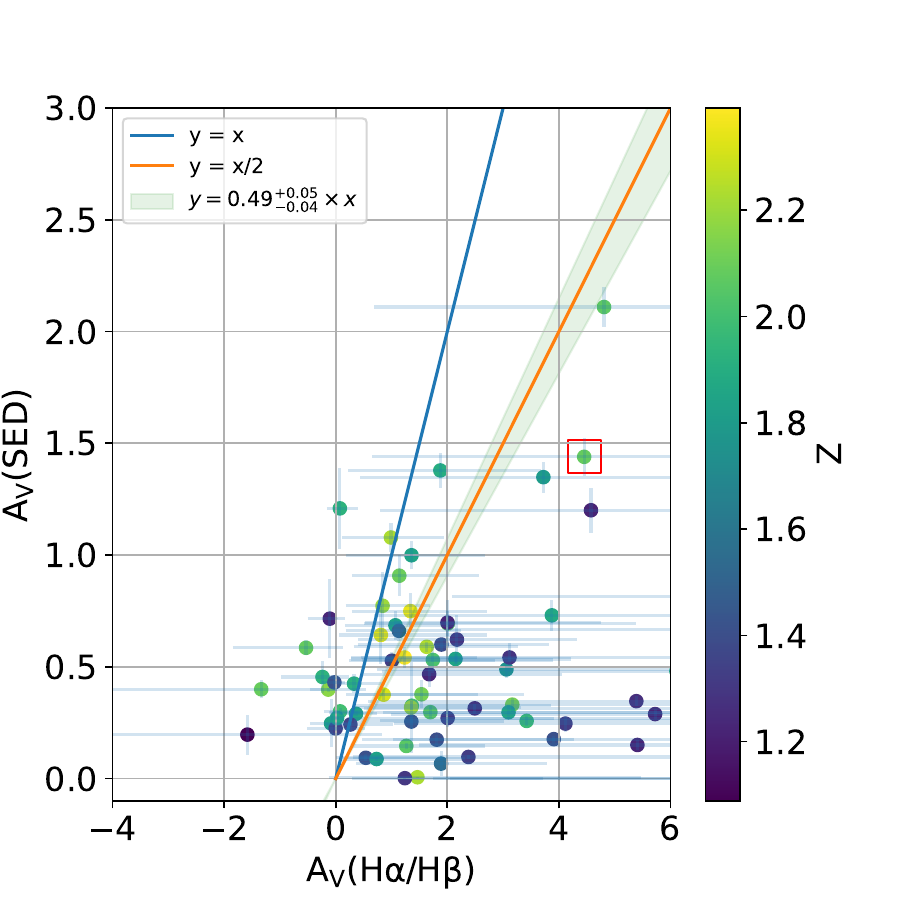}
\caption{Balmer decrement derived estimates of \AV, indicative of the dust content in individual star forming regions, versus photometry derived values of \AV\, indicative of the dust content of the galaxies as a whole, color coded by redshift. We observe that $\frac{A_V(SED)}{\AV(\Ha/\Hb)} = 0.49 ^{+0.05}_{-0.04}$.  X-ray AGN candidates from \citet{Luo17} are indicated using red squares.\label{Av_SED}}
\end{figure*}

Figure \ref{Av_SED} shows a comparison between \AV\ measured from the Balmer ratio with \AV\ inferred from SED fitting as explained in Section \ref{sec:SED}.  This comparison (plot and fitting) was restricted to only include derived values of \AV\ that were not significantly negative, and hence not significantly unphysical. The latter are expected to be representative of the dust content in the host galaxies as a whole, while the estimates derived using  emission line flux measurements are for the spectroscopically identified star forming regions. Further, Figure \ref{Av_SED} shows smaller values of \AV\ for star forming regions measured using the Balmer lines than for the host galaxies as a whole (using Prospector).  Quantitatively, the difference is $\frac{\AV(SED)}{\AV(\Ha/\Hb)} = 0.49^{+0.05}_{-0.04}$\ (when only considering the 19 objects with SNR$>2$\ Balmer decrement derived values of \AV).  A similar result was demonstrated by \citet{Papovich22}. In the rest of this paper, all of the emission line fluxes quoted are dust corrected based on the spectroscopic estimates of \AV\ (where available). For objects without usable Balmer lines, or those that clearly violate Case B, the Prospector derived values of \AV\ are used and corrected by the factor $0.49 ^{+0.05}_{-0.04}$.  Table \ref{tab:lines} lists the dust corrected line fluxes for the entire sample, as well as the values of \AV\ derived from the above analysis.

\newpage
\movetabledown=1.5in
\begin{rotatetable}
\begin{deluxetable}{ccccccccc}
\tabletypesize{\scriptsize}\tablecaption{Dust and contamination corrected emission lines and spectroscopically derived properties}\label{tab:lines} 
\tablehead{  \colhead{ID} & \colhead{z} & \colhead{H$\alpha$} & \colhead{H$\beta$} & \colhead{[OIII]$\lambda\lambda4960$} & \colhead{OIII$\lambda\lambda5008$} & \colhead{[OII]} & \colhead{Av} & \colhead{SFR(H$\alpha$)}\\ 
\colhead{ } & \colhead{ } & \colhead{$10^{-19}erg/s/cm^2$} & \colhead{$10^{-19}erg/s/cm^2$} & \colhead{$10^{-19}erg/s/cm^2$} & \colhead{$10^{-19}erg/s/cm^2$} & \colhead{$10^{-19}erg/s/cm^2$} & \colhead{mag} & \colhead{M$_\sun / yr$}} 
\startdata 
\multicolumn{9}{c}{AGN Candidates}\\ 
\hline 
 27914  & 2.08  & $492.53 ^{+7.59} _{-7.51}$ & $66.43 ^{+6.49} _{-6.66}$ & $39.00 ^{+6.55} _{-7.20}$ & $178.28 ^{+9.53} _{-9.00}$ & $199.36 ^{+13.36} _{-14.48}$ & $4.45 ^{+4.27} _{-3.80}$ & $80.01 ^{+925.57} _{-68.63}$\\ 
 32038  & 3.18  &   & $109.10 ^{+5.62} _{-5.87}$ & $121.77 ^{+5.25} _{-5.34}$ & $306.02 ^{+6.11} _{-6.06}$ & $14.55 ^{+3.47} _{-3.40}$ &   &  \\ 
\hline 
\multicolumn{9}{c}{Star Forming Galaxies}\\
\hline 
 33273  & 1.85  & $306.95 ^{+5.69} _{-5.70}$ & $109.43 ^{+3.41} _{-3.36}$ & $254.58 ^{+3.61} _{-3.59}$ & $635.79 ^{+4.34} _{-4.45}$ & $31.22 ^{+7.65} _{-7.50}$ & $0.02 ^{+0.13} _{-0.11}$ & $3.99 ^{+0.30} _{-0.50}$\\ 
 35875  & 2.23  & $1295.55 ^{+12.61} _{-13.00}$ & $392.48 ^{+7.57} _{-7.76}$ & $842.90 ^{+8.00} _{-7.91}$ & $1752.73 ^{+9.77} _{-9.49}$ & $633.01 ^{+16.36} _{-16.24}$ & $0.99 ^{+0.95} _{-0.87}$ & $38.36 ^{+16.91} _{-9.62}$\\ 
 26465  & 1.30  & $176.96 ^{+4.71} _{-4.62}$ & $107.06 ^{+8.45} _{-8.30}$ &   & $351.54 ^{+12.17} _{-12.45}$ &   & $-1.55 ^{+1.28} _{-1.52}$ & $1.24 ^{+0.12} _{-0.10}$\\ 
 27581  & 2.68  &   & $38.70 ^{+7.53} _{-7.00}$ & $48.99 ^{+6.37} _{-6.21}$ & $114.52 ^{+7.02} _{-7.03}$ & $69.83 ^{+5.25} _{-5.29}$ &   &  \\ 
 29659  & 3.32  &   & $15.30 ^{+4.87} _{-4.97}$ & $41.61 ^{+5.13} _{-4.81}$ & $120.45 ^{+6.50} _{-6.54}$ & $40.43 ^{+6.41} _{-5.87}$ &   &  \\ 
 33017  & 2.07  & $90.85 ^{+5.65} _{-5.60}$ & $39.15 ^{+3.42} _{-3.46}$ & $93.75 ^{+3.51} _{-3.40}$ & $225.49 ^{+4.21} _{-4.32}$ & $21.39 ^{+7.44} _{-7.10}$ & $-0.87 ^{+0.67} _{-1.37}$ & $3.51 ^{+1.19} _{-0.80}$\\ 
 34323  & 1.09  & $150.85 ^{+6.27} _{-5.93}$ & $115.98 ^{+26.64} _{-25.01}$ & $68.20 ^{+10.34} _{-12.06}$ & $228.41 ^{+14.64} _{-12.24}$ &   & $-2.61 ^{+2.01} _{-2.79}$ & $0.59 ^{+0.05} _{-0.04}$\\ 
 34389  & 1.09  & $48.31 ^{+7.62} _{-6.03}$ & $79.74 ^{+23.78} _{-23.18}$ & $38.24 ^{+15.68} _{-16.05}$ & $130.96 ^{+17.68} _{-17.07}$ &   & $-4.82 ^{+3.91} _{-4.91}$ & $0.17 ^{+0.03} _{-0.02}$\\ 
 35250  & 1.69  & $28.13 ^{+9.23} _{-8.94}$ &   &   &   &   &   & $1.13 ^{+0.95} _{-0.50}$\\ 
 36241  & 3.33  &   & $79.36 ^{+7.09} _{-6.83}$ & $191.44 ^{+7.29} _{-7.29}$ & $529.35 ^{+9.46} _{-9.35}$ & $148.07 ^{+8.15} _{-8.46}$ &   &  \\ 
 37266  & 3.42  &   & $46.95 ^{+5.21} _{-5.09}$ & $116.51 ^{+6.06} _{-5.78}$ & $284.90 ^{+8.53} _{-8.18}$ & $47.64 ^{+7.40} _{-7.19}$ &   &  \\ 
 38883  & 0.76  & $95.60 ^{+8.33} _{-7.96}$ &   &   &   &   &   & $0.27 ^{+0.08} _{-0.06}$\\ 
\enddata 
\tablecomments{The two objects detected in the X-ray \citep{Luo17} with fluxes $\ge$ 10$^{42}$ erg s$^{-1}$ are listed at the top part of this Table. Full table is available online}\end{deluxetable} 

\end{rotatetable}
\newpage

\subsection{Star Formation Rates}
\begin{figure*}
\center
\includegraphics[width=5in]{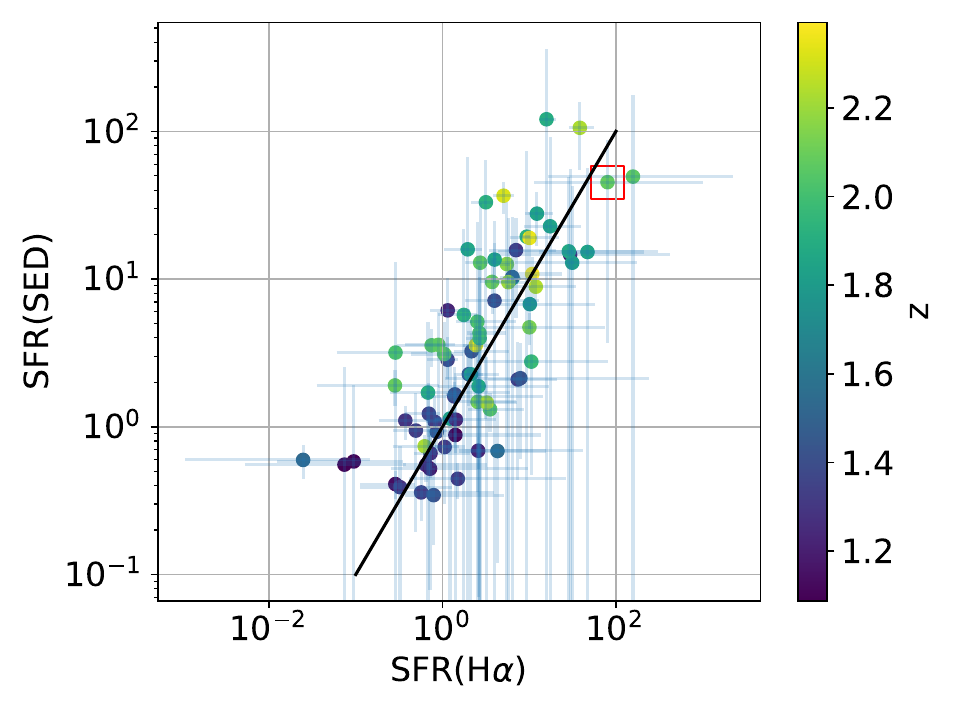}
\caption{Star formation rates as estimated using broad band photometry with Prospector versus our de-reddened, \NIIg\ decontaminated \Ha\ results, color coded by redshift. While in relatively good agreement, we note that the \NGDEEPA\ SFR estimates are significant better constrained than those derived using photometry measurements alone. X-ray AGN candidates from \citet{Luo17} are indicated using red squares.\label{SFR}}
\end{figure*}

The instantaneous star formation rates of galaxies in the sample were estimated directly from the dust corrected \Ha\ emission line fluxes \citep{Kennicutt12} listed in Table \ref{tab:lines}. The computed star formation rates are shown in the last column of Table \ref{tab:lines}.  As shown in Figure \ref{SFR}, the \NGDEEPA\ galaxies are typically forming stars at a rate of 0.1 to 1 ${\rm M_\sun / yr}$.
The star formation rates derived using Prospector with the photometric measurements (${\rm SFR_{SED}}$) are compared to the star formation rates derived directly from the WFSS. Figure \ref{SFR} shows these two estimates color coded by redshift. The two sets of estimates are consistent between the approaches. Figure \ref{SFR} shows that many values are roughly consistent (to within a factor $\simeq$ 2), however the Prospector estimated SFR appear to be significantly over-estimated at higher redshifts. . 

\begin{figure*}
\center
\includegraphics[width=6.5in]{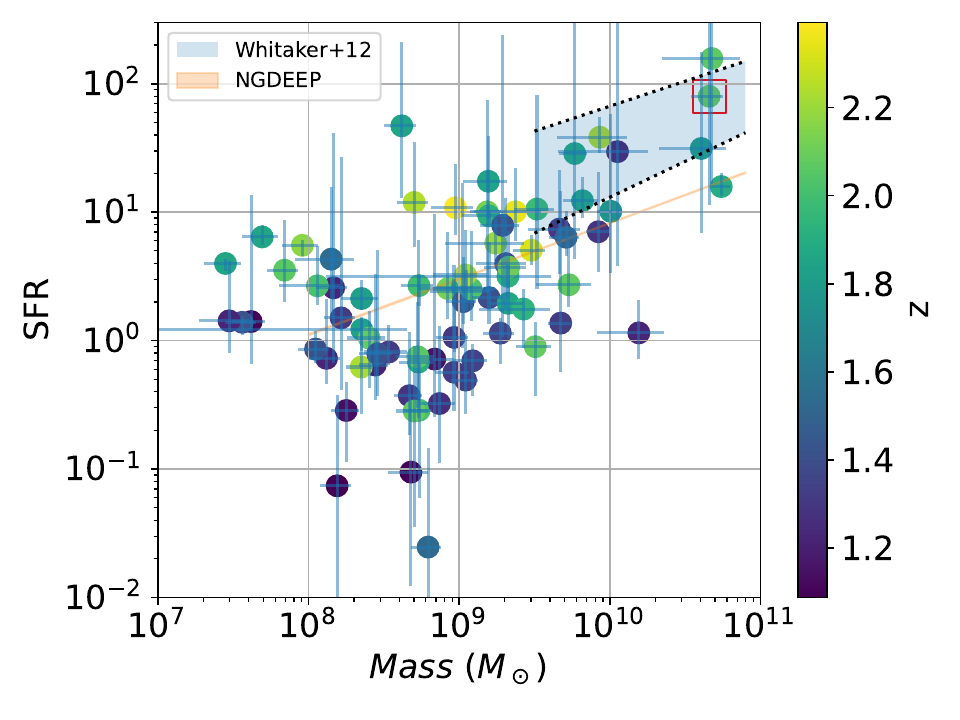}
\caption{The \NGDEEPA\ star formation rate estimates as a function of stellar mass. A linear fit and its 68\% confidence region for the \NGDEEPA\ sample (76 sources at $1.09 < z <  2.39$) is shown in orange.   We also show the relation from \citet{Whitaker12}, derived from photometric measurements of the NEWFIRM Medium-Band Survey, which is sensitive only to stellar masses of $10^{9.5} M_\sun$ and larger, and computed over the redshift range of our sample. \label{SFR_mass}}
\end{figure*}

Figure \ref{SFR_mass} shows the \citet{Whitaker12} SFR-Mass relations computed between the redshifts of 1.1 and 2.39 using photometric measurements from the NEWFIRM Medium-Band Survey (NMBS). This work was also limited to more massive objects with stellar masses greater than $10^{9.5}\ M_\sun$. As this figure shows, the \NGDEEPA\ extends down to lower SFR and smaller stellar masses due to greater sensitivity of emission line fluxes from \NGDEEPA. We find a slope in the mass to SFR relation with a linear fit of 
$0.37^{+0.007}_{-0.007}$ over the mass range of  $10^{7} < M_\sun <10^{11}$  using the \NGDEEPA\ sample. In comparison, the \citet{Whitaker12} relation predicts slopes between 0.39 and 0.55 over the redshift ranges shown and for masses $> 5 \times 10^9 \ M_\sun$.  While we derive SFR which appear  lower at any fixed redshift than those derived in the NMBS survey, both studies show an increase of SFR as a function of stellar mass and a  closer comparison is difficult as the NMBS results are based {\it solely} on the use of photometry from medium bandwidth filters, while \NGDEEPA\ uses measured emission lines. Further, NMBS did not include nebular emission in their analysis. 

\subsection{[OIII]/$H\beta$}
Another interesting diagnostic enabled by deep spectroscopy is the \OIIIg/\Hb\ or \OIII5008/\Hb\  ratio. This has long been used \citep{Baldwin81} to separate normal galaxies with HII star forming regions, planetary nebulae, from galaxies that are strongly photo-ionized, either by an active galactic nucleus or from ionizing shocks produced by massive young stars. As shown in \citet{Kewley13}, the  \OIII5008/\Hb\ ratio of star forming galaxies increases as a function of redshift, and is typically greater than 0.6 at $z>1.5$. \NGDEEPA\ allows us to examine the \OIII5008/\Hb\ ratio at higher redshifts, up to, and beyond Cosmic Noon ($z\approx2$). Nominally, one could use the R23 method \citep[e.g.][]{Pagel79,Pagel81} to estimate the gas metallicity of some of our sources, this method uses emission line fluxes over a wavelength range that spans 0.3 to 0.5 $\mu m$. R23 therefore is more sensitive to dust corrections, whereas the \OIIIg/\Hb\ or \OIII5008/\Hb\ are relatively close in wavelength and hence significantly less affected by variations in dust.

\begin{figure*}
\center
\includegraphics[width=6.5in]{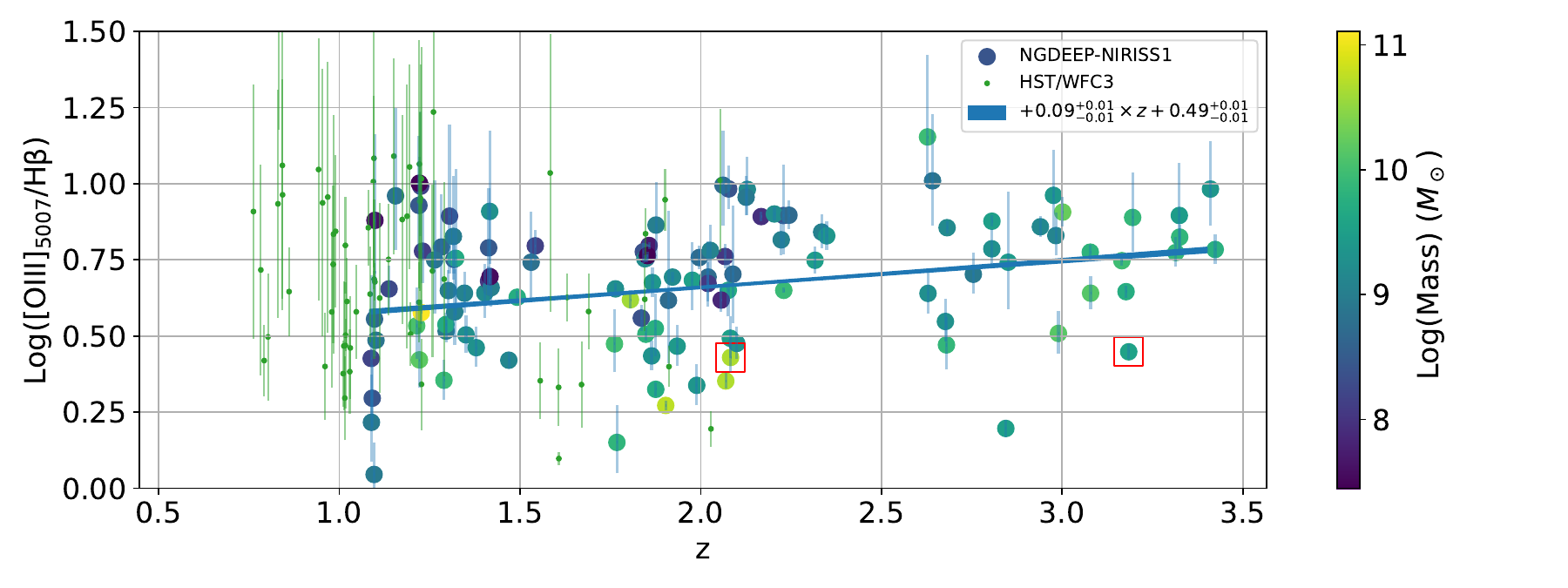}

\caption{\NGDEEPA\ \OIII5008/\Hb\ as a function of redshift shown, color coded by stellar mass. We show the distribution from the lower redshift, but noisier, HST/WFC3 G102 and G141 measurements in green. A linear fit to the \NGDEEPA\ data and its 68\% confidence region are shown in blue.  X-ray AGN candidates from \citet{Luo17} are indicated using red squares.\label{O3z}}
\end{figure*}

Figure \ref{O3z} shows the \NGDEEPA\ values of \OIII5008/\Hb\ with SNR$>2$\ as a function of redshift. Also plotted are the shallower measurements (also restricted to SNR $>$ 2) obtained from archival WFC3 G102 and G141 data for comparison  \citep[][Pirzkal et al., in prep]{Pirzkal18}. Although these cover a lower redshift range (0.8$<$ z $<$ 2), the range does overlap some with the JWST redshift range and they are consistent with the \NGDEEPA\ results.
Using the \NGDEEPA\ measurements, we detect a significant increase in the \OIII5008/\Hb\ line ratio as a function of redshift. Using this, a linear fit was made to the data. The fit (accounting for all errors), between $1<z<3.5$\ is: $ Log(\frac{[OIII]}{H\beta}) = 0.09^{+0.01}_{-0.01} \times z + 0.49^{+0.01}_{-0.01}$. This corresponds to an average increase of the [OIII] to $H\beta$\ line flux ratio of nearly 23\% per unit redshift (i.e. 1--2 Gyr).

\subsection{BPT}
Cross correlating the master source catalog to the \citet{Luo17} Chandra 7Ms X-ray catalog, four of the \NGDEEPA\ sources (IDs $\#$27914, 28858, 30372, and 32038) are detected in the X-ray by \citet{Luo17}. They have X-ray luminosities of $2.71 \times 10^{42}, 5.69\times 10^{41}, 6.15\times 10^{41}, 1.01\times 10^{44}$\ erg s$^{-1}$, respectively.  Sources 28858 and 32038 are identified as AGN dominated,  as their luminosities are above the nominal threshold of $\simeq$ 10$^{42}$ erg s$^{-1}$, while the other two sources are below this luminosity threshold and are characterized by \citet{Luo17} as star-forming dominated galaxies.

\begin{figure*}
\center
\includegraphics[width=7.5in]{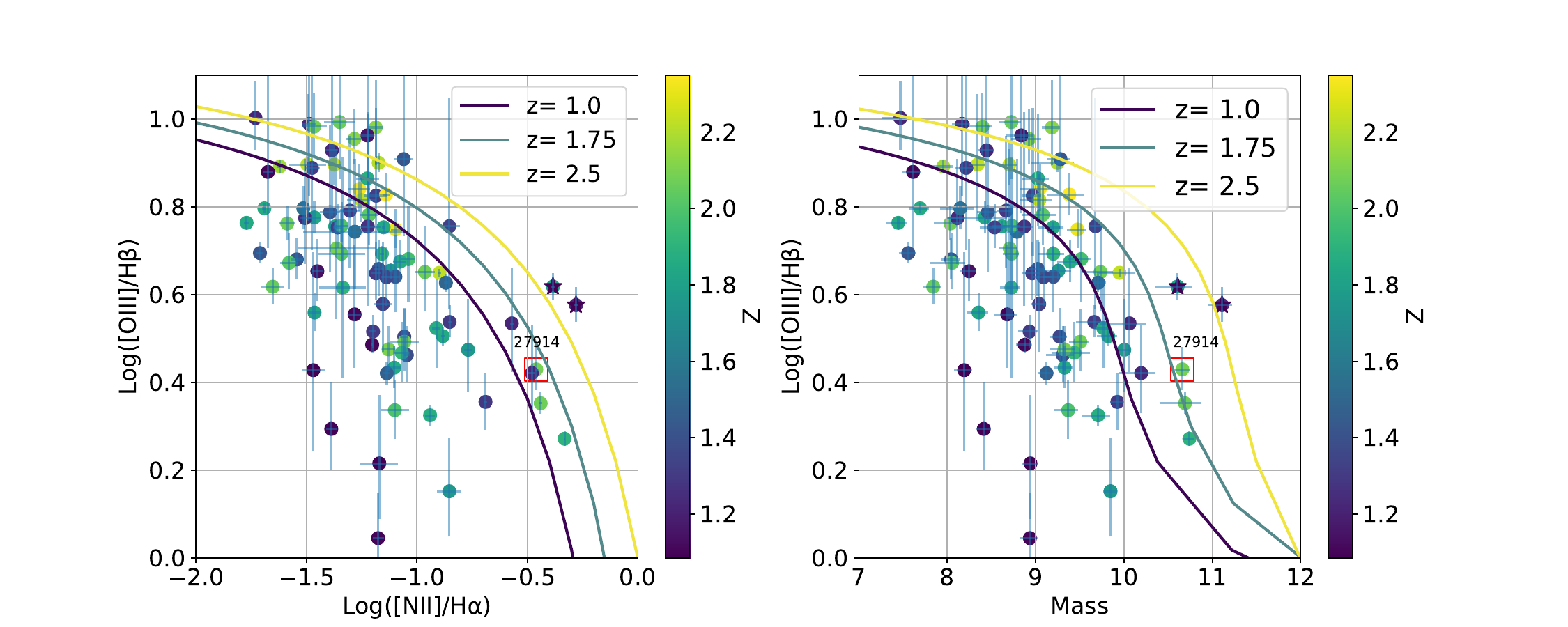}
\caption{\NGDEEPA\ BPT Diagram where the \citet{Faisst18} statistical relation between stellar mass and [NII]/H$\alpha$\ ratio is used (Left Panel), and  [OIII]/H$\beta$\ ratio shown as a function of stellar mass (Right Panel), for \NGDEEPA\ sources at $z<2.5$. 
Objects from \NGDEEPA\ which are a above the relation of  \citet{Kewley13} at 2$\sigma$\ significance, computed at their individual redshifts are plotted using a star symbol. X-ray AGN candidates from \citet{Luo17} are indicated using red squares.
The three colored lines are the \citet{Kewley13} computed  BPT lines separating AGN from SF dominated sources at the redshifts of 1.0, 1.75, and 2.5.}
\label{fig:BPT}
\end{figure*}

While individual, resolved measurements of the [NII] line fluxes for the \NGDEEPA\ galaxies are not possible due to the limitations of the NIRISS spectral resolution, we can use the statistically based \Ha/\NII6584-to-mass relation from \citet{Faisst18} (see Section \ref{[NII]}), combined with the stellar mass estimates derived from the SED fitting.  
Using the redshift dependent relation of \citet{Kewley13}:

\begin{equation}
Log([OIII]/H\beta) = \frac{0.61}{Log([NII]/H\alpha) - 0.02 - 0.1833 \times z)}+ 1.2 + 0.03 \times z
\end{equation}

while using Equation \ref{eq:faisst18} to estimate the corrected $Log([NII]/H\alpha$)\ from the stellar mass estimates, \NGDEEPA\ sources with measured  \OIII5008/\Hb\ line fluxes are plotted on a BPT diagram shown in Figure \ref{fig:BPT}. This figure shows both a typical BPT diagram (left panel) and and a version which uses stellar mass in place of $Log([NII]/H\alpha$) (right panel) and includes all sources for which there is stellar mass estimate. The right panel in Figure \ref{fig:BPT} is similar to the Mass Excitation (MEx) diagnostic developed in \cite{Juneau11}, which is applicable to sources out to z $\simeq$ 1 and \citet{Juneau14}, which extended the relation to z $\simeq$ 2.  The sources plotted in Figure \ref{fig:BPT} are limited to objects at $z<2.5$\, due to the redshift limitations of the computed lines from \citet{Kewley13}, as well as the redshift upper limits for the correction from \citet{Faisst18}.  

The two X-ray AGN candidates from \citet{Luo17} are denoted in several figures in this paper with a red square surrounding their data points.  ID$\#$ 27914 is plotted in Figure \ref{fig:BPT}. Two additional potential AGN candidates are shown in Fig \ref{fig:BPT}, plotted as stars (ID$\#$ 27037 and 28858).  They are $>$ 2$\sigma$ from their respective (redshift-dependent) BPT lines, but neither appear in \citet{Luo17} as X-ray sources, and to-date, no additional data from \NGDEEPA\ or other sources corroborate identifying them as AGN.  Several other sources appear above their respective BPT lines in Fig \ref{fig:BPT}. However, it should be noted that given their error bars, there still remains uncertainty in whether they are dominated by an AGN or starburst.  The addition of \NGDEEP\ Epoch 2 data should reduce the errors and may yield additional AGN candidates.

The second X-ray AGN candidate from \citet{Luo17}, ID$\#$ 32038 is at z $=$ 3.18, which places $\Ha$\ beyond the spectroscopic limits of the \NGDEEPA\ sample and obviates the use of a BPT diagram to test their AGN status. However, several emission lines within the NIRISS observing window further supports that this object is a strong AGN candidate.  \Hb\ and MgII emission lines are not only detected, but appear significantly broader than the instrumental resolution, especially when compared with other visible emission lines. The \Hb\ line is measured to have a FWHM of 92$\pm$ 3.4\AA\ for the average of the GR150R and GR150C grisms, while the average FWHM of the \OIII 5008 line is 46$\pm$ 0.94\AA (close to the nominal resolution of the NIRISS grisms). This corresponds to a broadening of 1191$\pm$ 57 km s$^{-1}$, for \Hb\, consistent with what would be expected from an AGN. For the MgII line, the broadened profile corresponds to 2259$\pm$ 24\ km s$^{-1}$. There is no shift in the observed wavelengths of the emission lines when observed with the GR150R and GR150C grisms.  Therefore, the physical source of the emission for these lines is located at the center of the source, as determined by SExtractor. 

\begin{figure*}
\center
\includegraphics[width=7.5in]{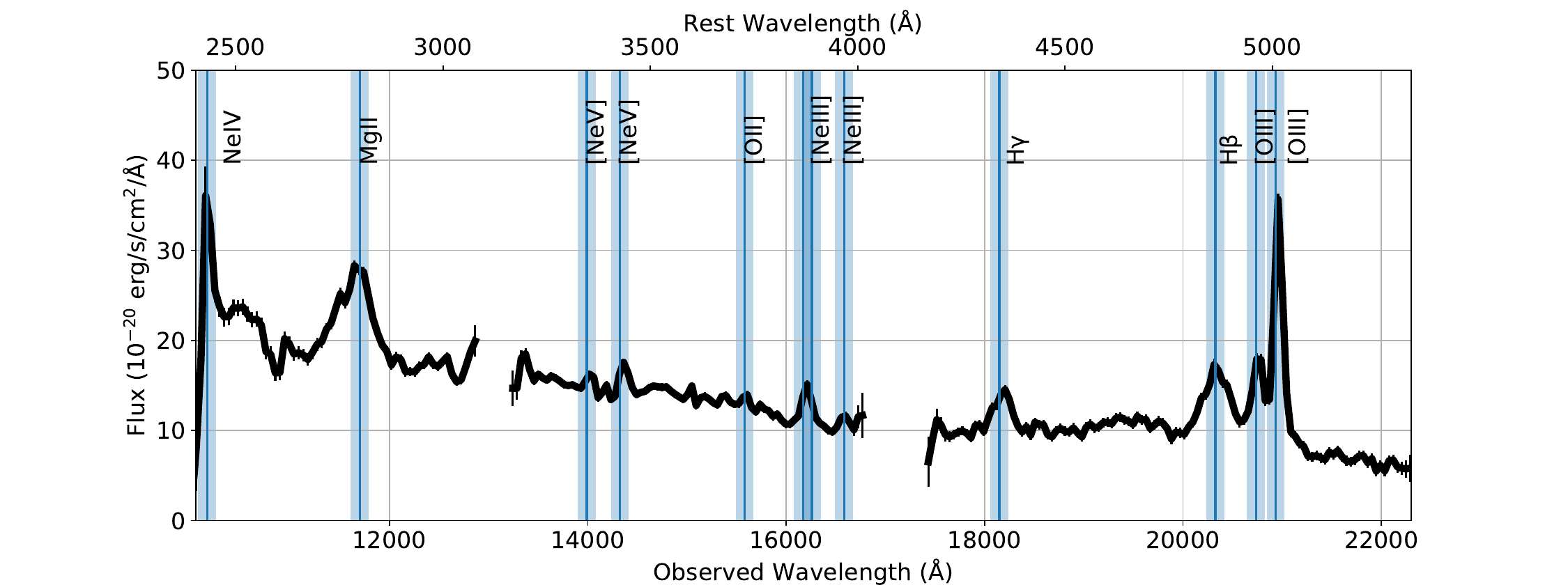}
\caption{\NGDEEPA\ spectrum of source 32038, an AGN dominated galaxy at z=3.18. This source is a known X-ray emitter \citep{Luo17} with an X-ray luminosity of 1.01 $\times$ 10$^{44}$ erg s$^{-1}$ and \NGDEEPA\ shows that its MgII and H$\beta$\ emission lines are significantly broadened. This object is not shown in Figure \ref{fig:BPT} as we do not have an \Ha\ flux for this source. We  measure line broadening of $\simeq$ 2200 km s$^{-1}$ and 1200 km s$^{-1}$ for MgII and H$\beta$\, respectively for this source, and derive a lower limit for the central black hole mass of $\approx 10^{8}\ M_\odot$.\label{AGN}}
\end{figure*}

The black hole masses can be estimated using the relation from \citet{Wang09}. Due to the absence of \Ha\ (which is redshifted beyond the sensitivity of NIRISS) and the absence of detectable \Hg\ emission, a dust correction cannot be applied, therefore the black hole mass estimates provided should be taken as a lower limit.  For \Hb\, using the measured F200W flux of 1.816$\pm$ 0.023$\mu$Jy, or a non-dust corrected luminosity of 9.91 $\pm$ 0.13 $\times$ 10$^{44}$ erg s$^{-1}$, leads to a black hole mass estimate of 1.71 $\pm$ 0.63 $\times$ 10$^{8}$ M$_{\sun}$. The MgII line luminosity of 8.94 $\pm$ 0.11 $\times$ 10$^{44}$ erg s$^{-1}$ and an F115W flux of 964$\pm$24 $\mu$Jy yields a black hole mass estimate of 1.38 $\pm$ 1.02 $\times$ 10$^{8}$ M$_{\sun}$. A full spectrum of this source, combining the GR150R and GR150C spectra, is shown in Figure \ref{AGN}.

\section{Conclusion \& Discussion}
This work demonstrates a first look at the extracted NIRISS WFSS spectra obtained as part of the \NGDEEP\ project.  Although only half of the allocated observations have been obtained, the results confirm the power of JWST in conjunction with NIRISS WFSS to probe the characteristics of emission line galaxies out to moderate redshifts, including leveraging the improved spectral resolution which allows for the detection of AGN and obtaining mass estimates of SMBHs at z $>$ 3.  The paper presents two distinct sets of results. The first, presents an analysis and demonstration of the need for careful extraction, calibration, and analysis of NIRISS WFSS data as outlined in \ref{WFSS Extraction and Calibration} and with additional details provided in Appendix \ref{calib}.   These results show that the improved calibrations can obtain an accuracy of better than 0.25 pixels for the geometric trace and 10$\AA$ or better for the wavelength calibration (equivalent to one-quarter of a resolution element, on average) regardless of location of the dispersed objects within the field of view.  The standard calibration files produce errors 2-3 pixels along the trace, and $\simeq$ 90-120$\AA$ in wavelength-space.  Further, this paper establishes a methodology for using these improved calibrations to avoid potential pitfalls when dealing with resolved or partially resolved objects by taking into account the inherent differences in the properties of the GR150C and GR150R grisms. As shown in Section \ref{Results}, a number of the scientific results, such as detecting individual star-forming regions within galaxies at z $>$ 1, or obtaining direct mass estimates for a SMBH at z $>$ 3, could not be achieved without the factor of 5-10$\times$ improvement from these new calibrations and methods. Appendix \ref{calib} further quantitatively demonstrates these improvements. As a service to the community, these calibration files are made publicly available and it is hoped that the details provided here will be useful for improving the data products produced from previous and future NIRISS WFSS observations.
 
The second set of results presented in this paper focus on the science which can be achieved with the NIRISS WFSS observations.  The improved spectral and spatial resolution obtained from the G150R and G150C grisms have made it possible to detect multiple emission lines not only from a single galaxy, but from multiple positions within galaxies.  The WFSS observations have been used to: characterize star-formation rates, estimate dust corrections using the Balmer decrement, demonstrate a redshift evolution in the ratio of \OIII5008/\Hb\, compare the correlation of SFR with stellar mass over redshift between \NGDEEPA\ other surveys, and ascertain whether the emission line galaxies in the sample are star-formation or AGN dominated using BPT and MEx diagnostics. The last result demonstrates a particularly useful survey and diagnostic power of the NIRISS WFSS data, namely that the instrument has sufficient spectral resolution to detect the presence of AGN and estimate the masses of SMBHs at sensitivies and redshifts beyond what has previously been possible.  Further, the paper has demonstrated the power of WFSS observations relative to using photometry alone to constrain the physical parameters of galaxies.

While a number of results shown in the paper have either been consistent with similar works at lower redshift, or at least demonstrated similar correlations, but with scale and depth differences (e.g. SFR vs stellar mass), one interesting result raises questions about the efficacy of certain assumptions.  As noted in \ref{sec:Av}, the use of the Balmer decrement to constrain and correct for the presence of dust has produced a significant number of sources which appear to violate Case B recombination.  Some first hints of this were seen in \cite{Matharu23}.  While there exist a number of possible explanations for such violations, one possibility, which has been raised in other surveys such as MOSDEF \citep{Shapley22}, early release science results from NIRISS \citep{Matharu23} etc, is poor signal to noise or significant errors in measuring the fluxes of \Ha\, \Hb\, etc.  Although \NGDEEP\ has only obtained half of its allocated observations, the signal to noise of objects which violate Case B are significantly high (e.g.  SNR$>$10-20).  For example, Source 36292, a $z=2.17$ galaxy with a Prospector estimated stellar age of a few million years and a mass of $10^8  M_\odot$\, for example, has \Hag, \Hb, and and \Hg\ emission lines with SNR of 54, 30, and 19, respectively. These lines are well detected in both GR150R and GR150C and yet result in a ratio of \Ha / \Hb\ of $2.67\pm0.1$ and \Hg / \Hb\ of $0.65\pm0.04$.

Another explanation presented in previous works, is to apply an absorption correction to the values of \Ha\, \Hb\, \Hg\, etc.  The assumption is that in addition to the star-forming regions, there exists a significant (luminosity weighted) stellar population which exhibit Balmer absorption lines.  When viewed along the line of sight, these absorption lines contaminate the measured emission lines.  Since the spectral resolution of the NIRISS observations are either too poor to detect the absorption lines, or the absorption lines are too faint to be detected relative to the already faint continuum, a emperical correction \citep[e.g.][]{Groves12} should be applied to the measured values.  It is usually assumed that Balmer absorption should remain very low for stellar population that are less than 1 Gyr old. Yet, galaxies in the \NGDEEPA\ sample typically have stellar ages that are only a few hundred million years (see \ref{HaHb} and \ref{HgHb}, Source 36292 as noted above).

This suggests that the cause of the violations is more likely to be related to the physical nature of the galaxies themselves.  We may be witnessing scenarios in which neither Case A, nor Case B are appropriate to use for dust corrections.  Modified versions of Case A and B, namely Case C or Case D \citep{Baker38, Ferland99, Luridiana09, PrivScarlatta}, in which the assumption is that neither all of Lyman $\alpha$ photons escape nor  are all  absorbed.  Case C and D suggest that HII regions may contain non-uniformly distributed dust which further scatters and absorbs photons.  Such an analysis is deferred to subsequent papers, but we raise the possibility that Case C or Case D may be more appropriate for galaxies with formation histories unlike those in the local Universe.  The second epoch of \NGDEEP\ with the resulting deeper data, and addition of two more position angles should help examine in more details in a follow up paper. Full forward modeling 2D maps will allow us to determine whether there are significant morphological issues in the host galaxies that affect how we currently measure our emission line fluxes.

\software{Astropy \citep{Astropy13, Astropy18, Astropy22},  
        Dynesty \citep{dynesty},
          Source Extractor \citep{Bertin96}
          }


\newpage

\section*{Acknowledgement}
This work is based on observations made with the NASA/ESA/CSA James Webb Space Telescope. The data were obtained from the Mikulski Archive for Space Telescopes at the Space Telescope Science Institute, which is operated by the Association of Universities for Research in Astronomy, Inc., under NASA contract NAS 5-03127 for JWST. These observations are associated with program 02079.
Support for program 02079 was provided by NASA through a grant from the Space Telescope Science Institute, which is operated by the Association of Universities for Research in Astronomy, Inc., under NASA contract NAS 5–26555.\\
The authors would like to extend their thanks to the referee and would also like to thank Karl Gordon and Claudia Scarlata for the helpful guidance and discussions regarding hydrogen recombination cases and dust extinction laws.
\facility{JWST (NIRISS), HST (WFC3/IR)}\\
The data described here may be obtained from the MAST archive at \dataset[doi:10.17909/02wx-6j29]{http://dx.doi.org/10.17909/02wx-6j29}.

\appendix
\section{JWST NIRISS \NGDEEPA\ Calibration} \label{calib}
At the time of the  \NGDEEPA\ observations, using the available STScI JWST NIRISS WFSS calibration (which we refer to as the {\em default} calibration in the rest of this paper), the location of the traces were typically 1-2 pixels off, limiting our ability to accurately model and subtract WFSS contamination. No field dependence to the shape of the traces were calibrated and only the center most part of the field of view was calibrated. The wavelength solution was determined at the center of the field and did not include any field dependence. The extractions based on the {\em default} calibration products therefore did not allow us to properly combine R and G grisms. This is something that can only be done if both grisms have each been calibrated accurately over the entire field of view, and if they have each been calibrated consistently. However, Commissioning and Cycle 1 observations of stellar calibrators (PID 01089 using P330-E and WD1657+343 for trace geometry and flux calibrations; and PID 01510 using LMC PN58 for wavelength calibration) were obtained. During the course of these observations, the calibrator objects were placed at different positions in the detector which allowed for the proper derivation of the field dependence of the NIRISS WFSS grisms dispersions. 

In this Appendix, we describe the \NGDEEP\  spectral traces and wavelength calibration of the dispersion of the NIRISS R and C grisms when using the F115W, F150W, and F200W filters. We also show the accuracy of the trace geometry, and wavelength  calibration we derived. This work follows the methodology used in the past to calibrate the Hubble Space Telescope Advanced Camera for Surveys (ACS), and the Wide Field Camera 3 (WFC3). Those previous efforts were documented in the form of Instrument Science Reports (ISRs, e.g. \cite{Pirzkal16,Pirzkal17c}). In this work, we processed all uncalibrated JWST data using the STScI Pipeline Stage 1 and Stage 2, to produce {\em RATE} files (in units of DN/s) that were: bias and dark subtracted;  on-the-ramp fitted; flat-fielded using the appropriate imaging filter;  and fully populated with a world coordinate system. The pipeline Stage 2 was also used to produced {\em CAL} imaging files which were combined into deep cosmic ray free mosaics using Stage 3 of the STScI JWST pipeline. It should be noted that some of the calibrators wer saturated in these exposures, making it impossible to measure their positions accurately (i.e. $<0.1$\ pixel accuracy).  Instead, we measured the accurate positions of other non saturated stars in the fields, matched those to the expected positions derived using the GAIA DR3 catalog, and computed an affine transformation between the GAIA DR3 and each of our mosaic reference frames. The position of the saturated calibrator were then computed by applying the derived affine transformation to the GAIA DR3 coordinates of these sources. This approach worked well and provided better than 0.1 pixel accuracy.

\subsection{Trace Geometry}
\begin{figure*}
\center
\includegraphics[width=6.5in]{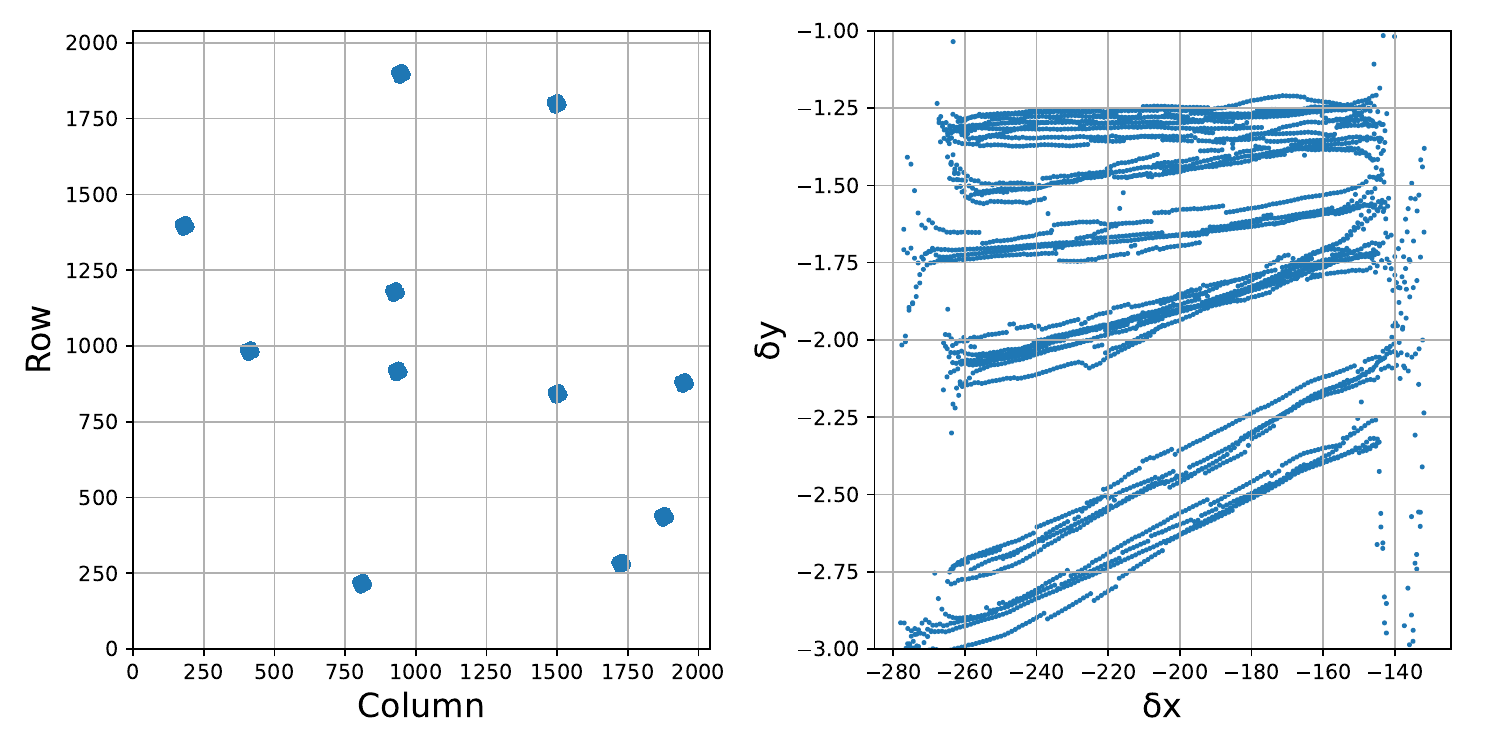}
\caption{\NGDEEP NIRISS Trace measurements using PID 01089 observations using the GRISM R crossed with the F200W imaging filter. The shape of traces were measured across the entire detector (Left Panel) and variation in the geometry of the spectral traces are readily visible in the Right Panel.\label{WFSS_geo}}
\end{figure*}

\begin{figure*}
\center
\includegraphics[width=6.5in]{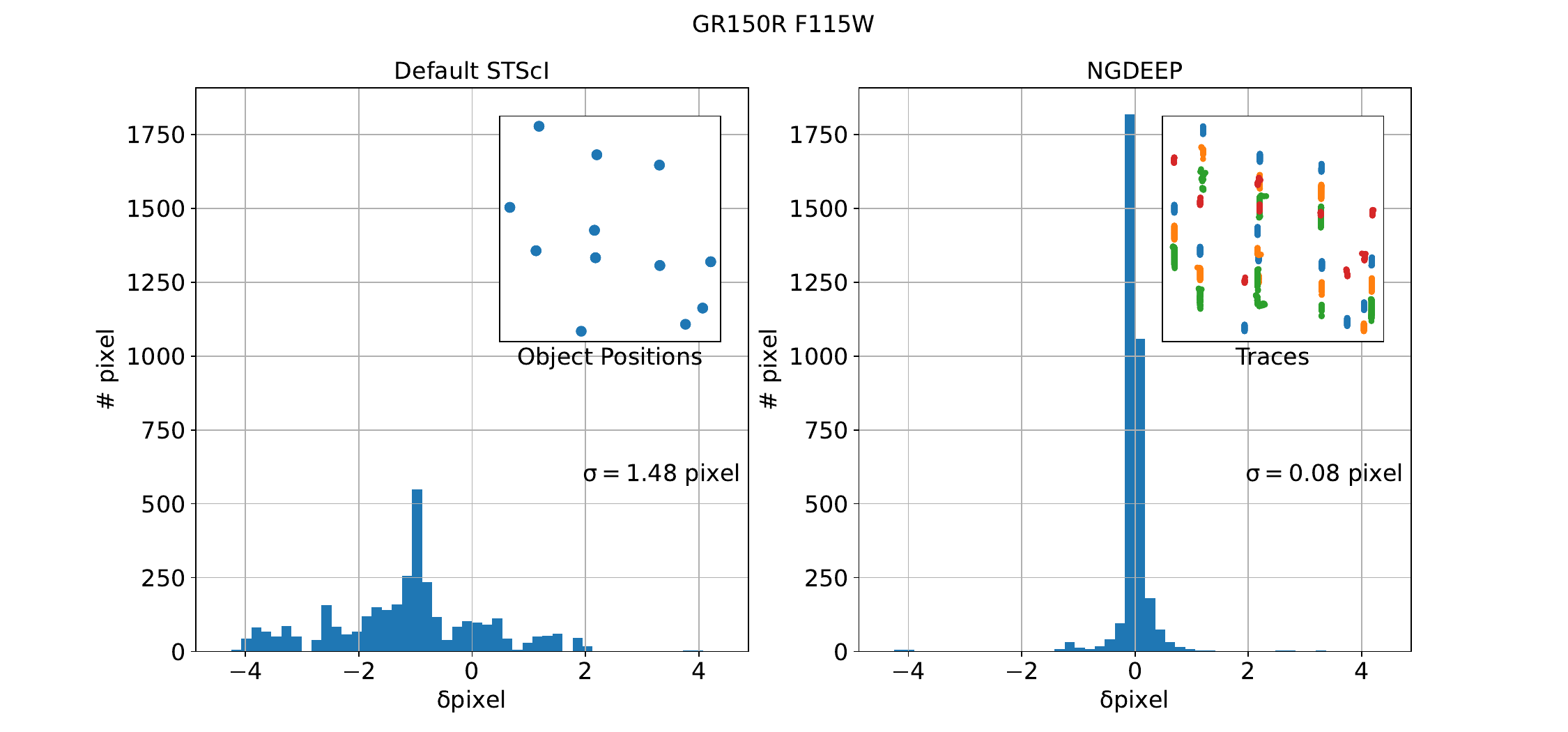}
\caption{Comparison of the official NIRISS calibration and the \NGDEEP\ calibration we adopted in this work when using the GR150R grism with the F115W filter. The left panel shows the histogram of the difference between measured and predicted location of the +1,+2,+3,-1 spectral orders. The inset of the Left Panel shows where calibration sources were located within the field of view of the instrument. The Right Panel shows the same histogram but computed using our \NGDEEP\ calibration. The inset of the Right Panel shows where the +1,+2,+3,-1 order traces were located on the detector.
.\label{geo1}}
\end{figure*}

\begin{figure*}
\center
\includegraphics[width=6.5in]{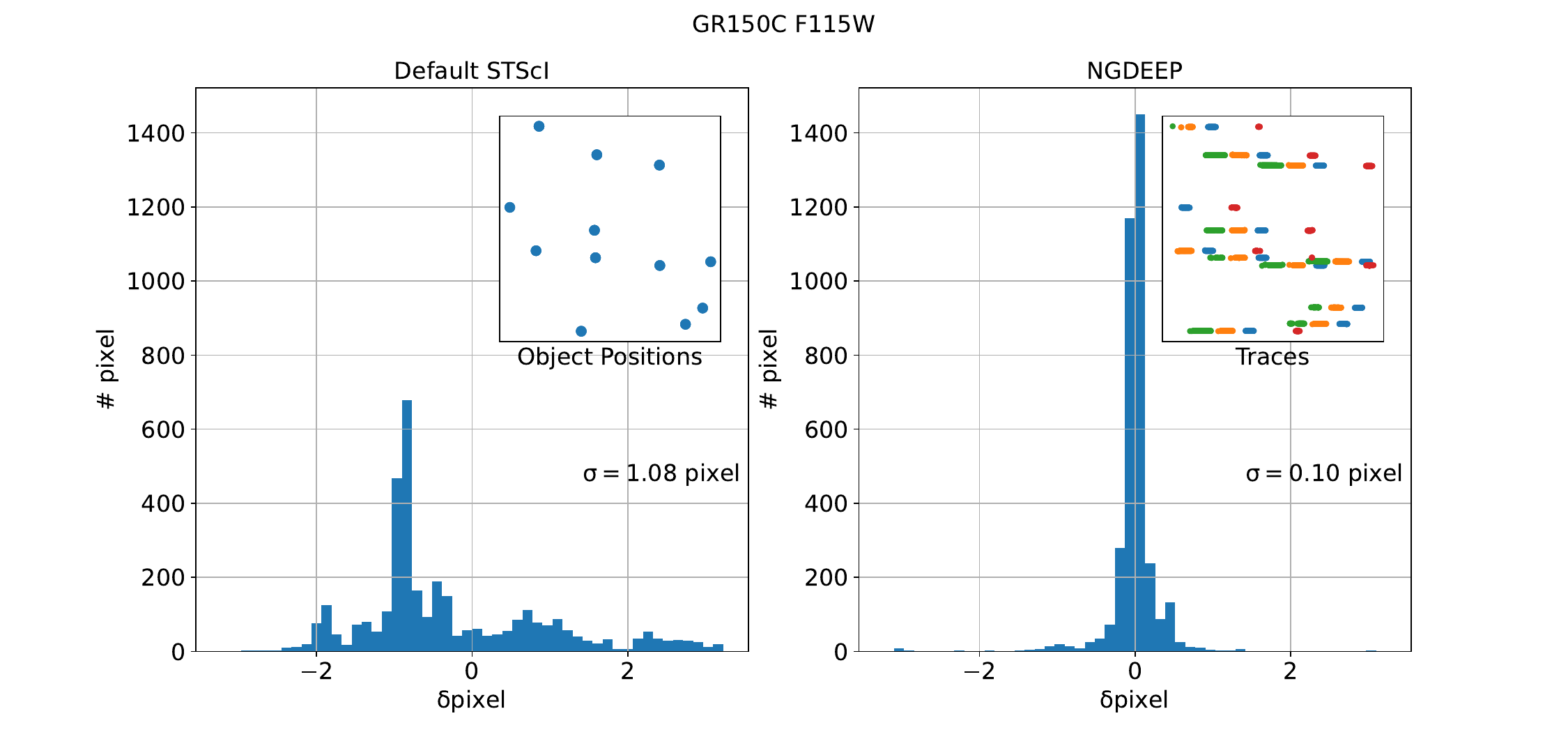}
\caption{Same as Figure \ref{geo1} but for the GR150C grism with the F115W filter.\label{geo2}}
\end{figure*}

\begin{figure*}
\center
\includegraphics[width=6.5in]{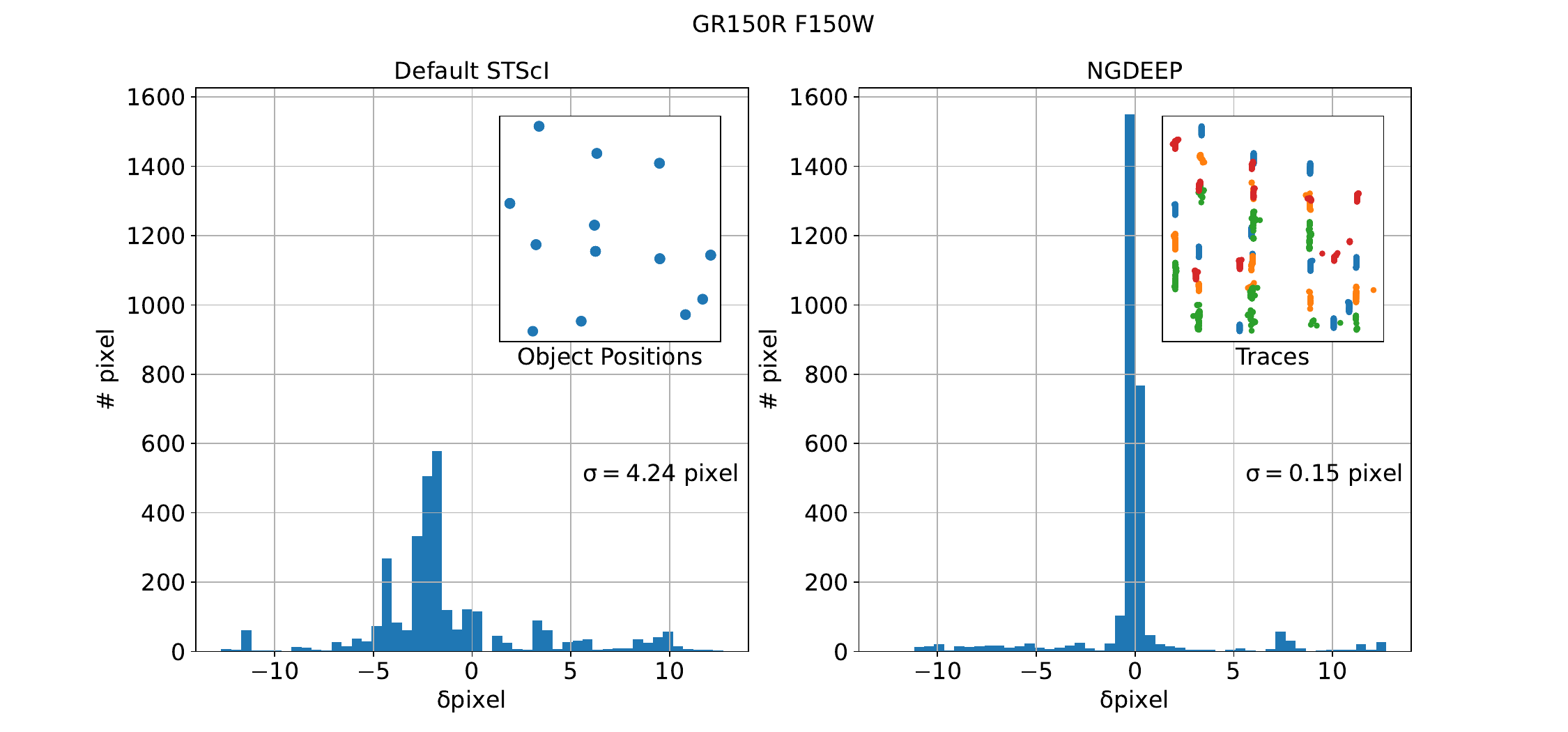}
\caption{Same as Figure \ref{geo1} but for the GR150R grism with the F150W filter.\label{geo3}}
\end{figure*}

\begin{figure*}
\center
\includegraphics[width=6.5in]{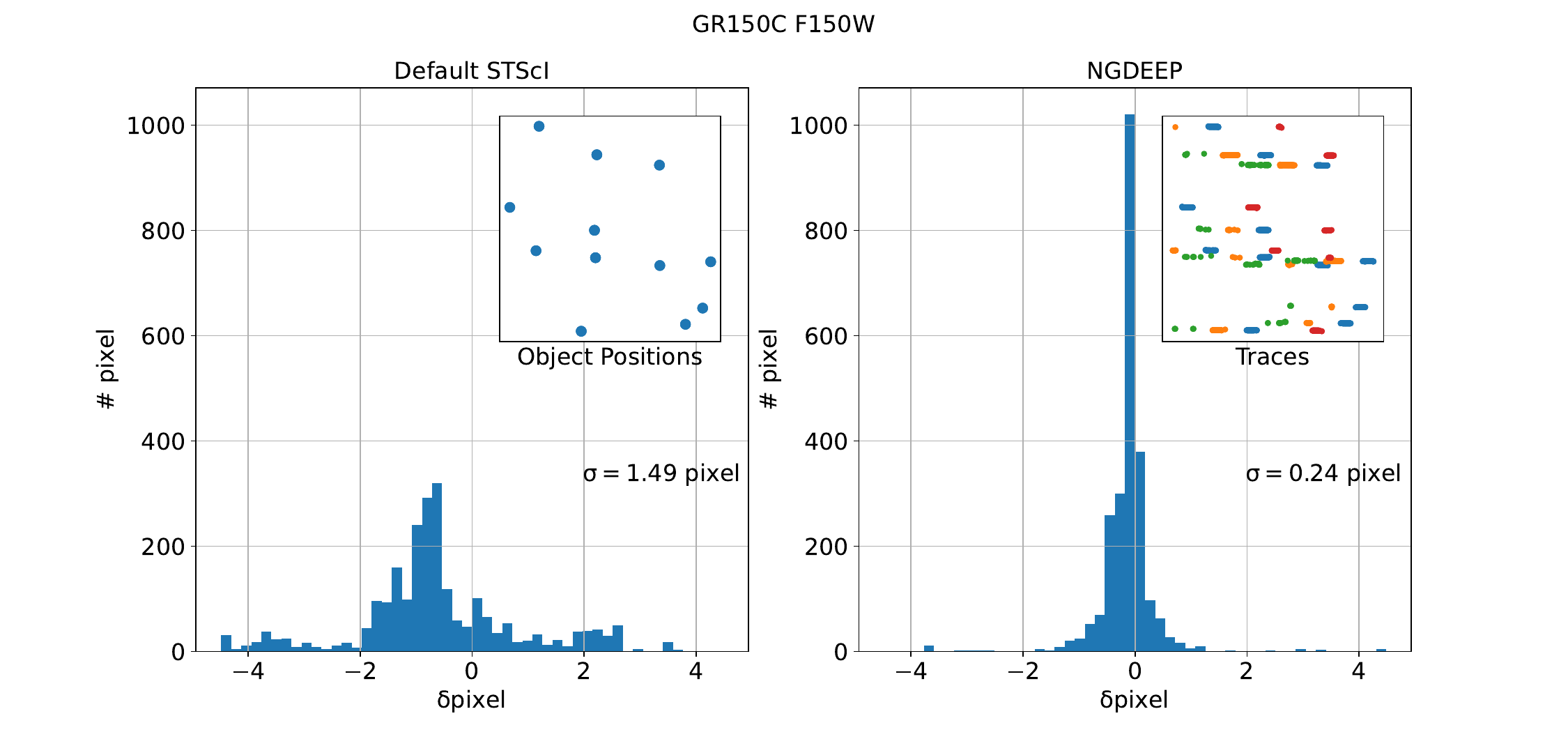}
\caption{Same as Figure \ref{geo1} but for the GR150C grism with the F150W filter.\label{geo4}}
\end{figure*}

\begin{figure*}
\center
\includegraphics[width=6.5in]{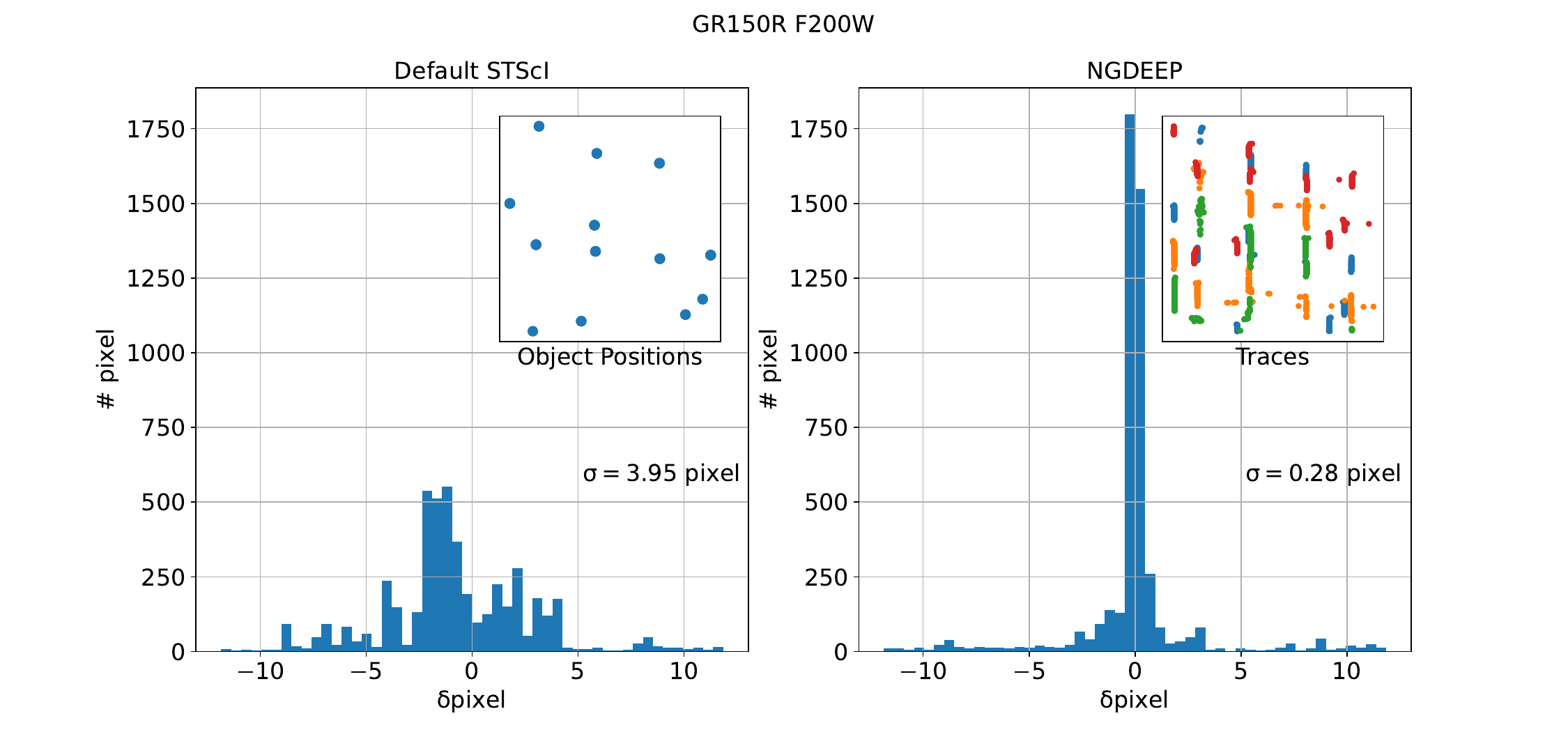}
\caption{Same as Figure \ref{geo1} but for the GR150R grism with the F200W filter.\label{geo5}}
\end{figure*}

\begin{figure*}
\center
\includegraphics[width=6.5in]{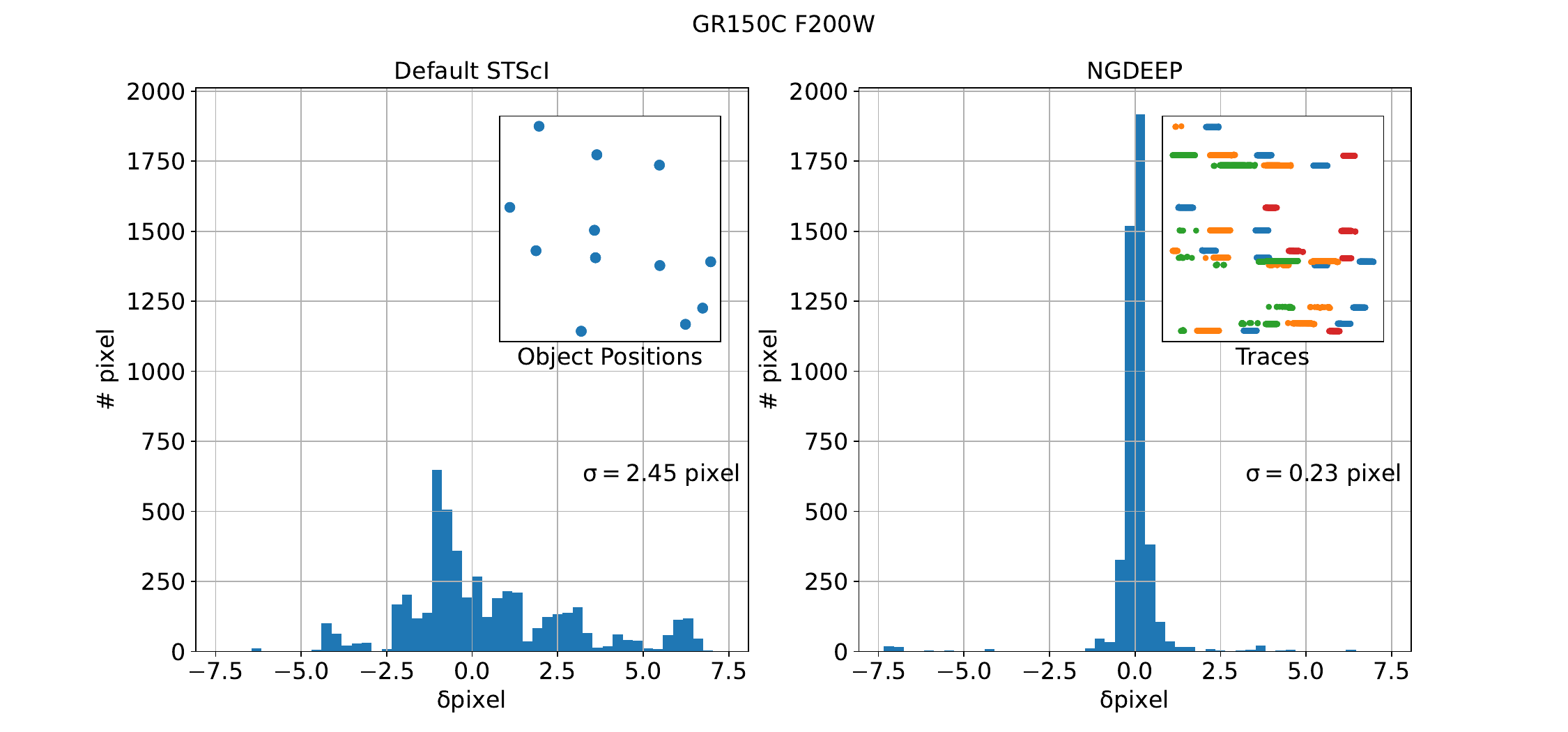}
\caption{Same as Figure \ref{geo1} but for the GR150C grism with the F200W filter.\label{geo6}}
\end{figure*}

As for previous WFSS mode (e.g. HST/ACS. HST/WFC3. HST/UVIS), all characteristics of the dispersing element are determined with respect to the position of the light emitting object in the field of view, in pixel coordinates ($x_0,y_0$).
For each combination of grism and cross filter, and using data from PID 01089, we were typically able to identify a dozen stars across the detector in the associated direct imaging, each dithered using a 4 point small dither pattern. After identifying the appropriate spectra order in the associated WFSS data, we proceeded to fit a Gaussian profile plus a linear continuum background in the cross dispersion direction while doing this along the entire length of the spectral order. As shown in Figure \ref{WFSS_geo}, the NIRISS WFSS traces have a relatively strong field dependence to both the cross dispersion offset between the source location ($x_0,y_0$) and the average cross dispersion position of the trace on the detector. The variability of the later is also shown in the right Panel of Figure \ref{WFSS_geo}.
Following the methodology described in \citet{Pirzkal17c}, we found that a second order polynomial with a 2nd order field dependence modelled the observation well. We performed the calibration for all 6 combinations of grisms (GR150R, GR150C) and filters (F115W, F150W, F200W) used by \NGDEEP. While the residuals from the {\em default} reference files lead to large differences between predicted and measured positions of the +1, +2, +3, and -1 orders which are on the order of several pixels, our \NGDEEP\ calibration  predicts the location of the traces to within a fraction of a pixel (Typically $<0.25$\ pixel, all over the detector. Figures \ref{geo1} to \ref{geo6} show the increased accuracy of the \NGDEEP\ trace calibration compared to what was officially available. Table \ref{tab:geo} summarizes the average difference between measured and predicted location of the traces when using the {\em default} calibration and the \NGDEEP\ calibration. Note that there is a known dependence on the tilt of the dispersed spectra and the filter wheel position (FWCPOS) and the NGDEEP calibration products accounted for this.

\begin{deluxetable}{cccc} 
\tablewidth{0pt} 
\tablecaption{\NGDEEP\ Calibration of Trace Geometry Residuals\label{tab:geo}} 
\tablehead{\colhead{GRISM} & \colhead{Filter} & \colhead{Default} & \colhead{\NGDEEP} \\
 &  & ${\rm Pixel}$ & ${\rm Pixel}$}  
\startdata 
GR150R & F115W  & 1.48 & 0.08  \\
GR150R & F150W  & 4.24 & 0.15  \\
GR150R & F200W  & 3.95 & 0.28  \\
GR150C & F115W  & 1.08 & 0.10  \\
GR150C & F150W  & 1.49 & 0.24  \\
GR150C & F200W  & 2.45 & 0.23  \\
\enddata 

\end{deluxetable}

\subsection{Wavelength Calibration}
\begin{figure*}
\center
\includegraphics[width=6.25in]{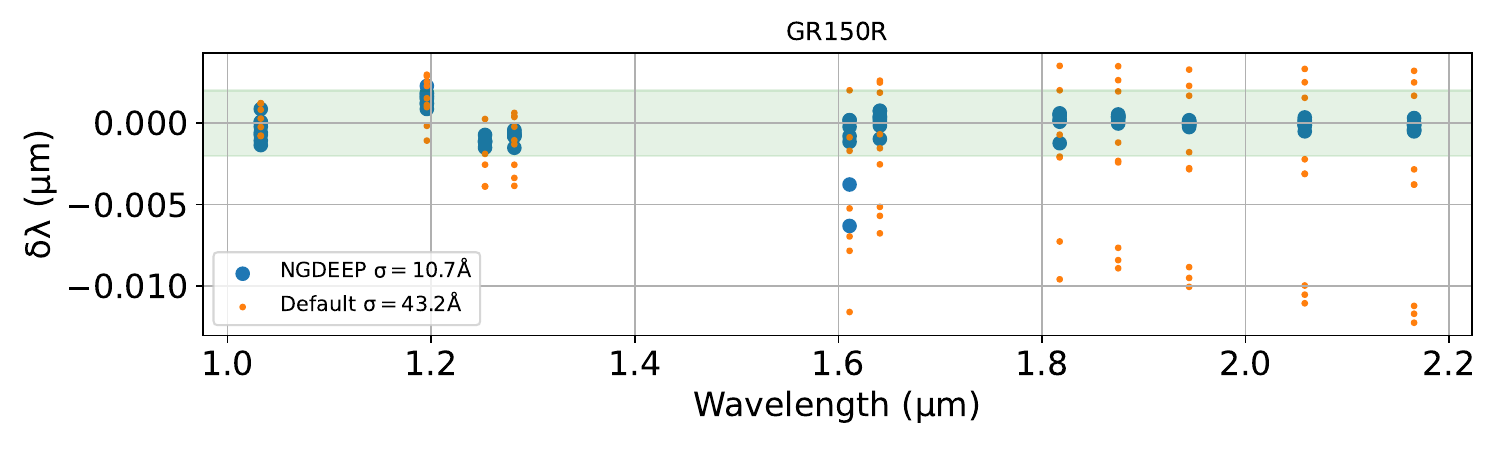}
\caption{Difference in computed emission line wavelengths in observations of PN58 using the \NGDEEP\ wavelength solution (blue) and the default available calibration (orange) when using the GR150R grism.
\label{wav1}}
\end{figure*}

\begin{figure*}
\center
\includegraphics[width=6.25in]{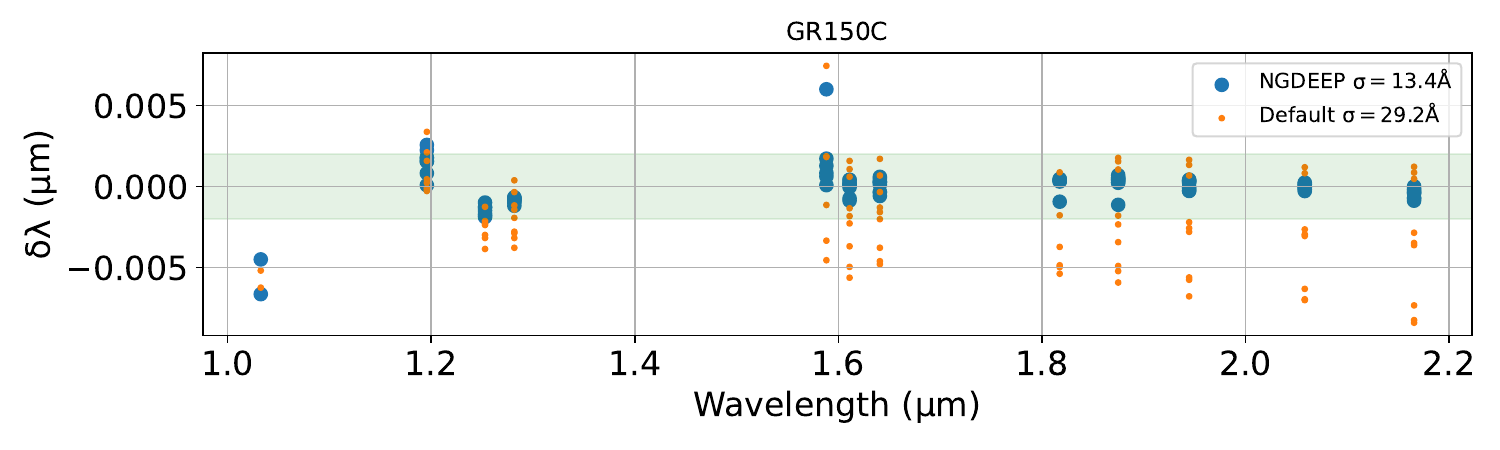}
\caption{Same as Figure \ref{wav1} but for the GR150C grism \label{wav2}}
\end{figure*}

NIRISS observed a wavelength calibrator (LMC PN58) at 9 positions across the detector, covering a large part of the field of view. We extracted the NIRISS spectra of the target and proceeded to measure the location of the emission line along the spectral traces. A global field dependent solution can be determined by comparing the observed positions along the traces to the fiducial wavelengths of the emission lines. As the NIRISS spectra have a limited wavelength range, a single wavelength solution was fit along the three cross filters used, covering the weavelength range of $1<\mu m<2.2$. The fiducial wavelengths of the lines were obtained using NIRSPEC MOS observations of the target.  A 1st order polynomial with a 2nd order field dependence was fit.  This type of polynomial was found to be optimal, as it allowed us to derive solutions that reduced the residuals of the fit to a small fraction of a pixel, typically better than $20\AA$\ or half of a pixel. The choice of the order was predicated on balancing minimizing residuals, while avoiding over fitting the data with extrapolated solutions. Over fitting the data with higher order would cause the fits to strongly diverge near the edges of the field of view. This is in contrast to the {\em default} calibration products which, even when extracting these spectra using the improved trace calibration described in the previous section, showed  progressively worse residuals.   At longer wavelengths these became on the order of several pixels/resolution elements or more ($\approx 20 -- 120\AA$). Figures \ref{wav1} and \ref{wav2}  show the differences between expected and predicted wavelengths along the traces of the R and G grisms. In summary, the \NGDEEP\ wavelength calibration of the NIRISS grisms is typically within $\approx 10\AA$ which is approximately a quarter of a resolution element and therefore on a similar scale as the trace calibration described in the previous section. There is a significant amount of field dependence to the wavelength calibration as shown by the large spread of errors when using the {\em default} NIRISS WFSS calibration.

\bibliographystyle{apj}
\bibliography{pirzkal}

\end{document}